\documentclass[epj]{svjour}
\usepackage{amssymb,amsmath,latexsym,fancyhdr,bbold,epsfig,graphicx}
\usepackage{epic,eepic}
\usepackage[latin1]{inputenc}
\usepackage[squaren]{SIunits}

\renewcommand{\vec}[1]{\boldsymbol{#1}}
\newcommand{\mat}[1]{\boldsymbol{\sf #1}}

\renewcommand{\d}{\text{d}}
\renewcommand{\@}{\partial}
\newcommand{\oo}{\infty}

\newcommand{\m}[1]{\left\langle #1 \right\rangle}
\begin{document}
\title{Motor-Driven Bacterial Flagella and Buckling Instabilities}
\author{Reinhard Vogel \inst{1} \and Holger Stark\inst{1}
}                     
%
%
\institute{Institute for Theoretical Physics , TU Berlin \and Institute for Theoretical Physics , TU Berlin}
\date{Received: date / Revised version: date}
%
\abstract{
Many types of bacteria swim by rotating a bundle of helical filaments
also called flagella. Each filament is driven by a rotary motor and
a very flexible hook transmits the motor torque to the filament.
We model it by discretizing Kirchhoff's elastic-rod 
theory and develop a coarse-grained approach for driving the helical 
filament by a motor torque. A rotating flagellum generates a thrust
force, which pushes the cell body forward and which increases with the 
motor torque. We fix the rotating flagellum in space and show that it
buckles under the thrust force at a critical motor torque. 
Buckling becomes visible as a supercritical Hopf bifurcation in the thrust 
force. A second buckling transition occurs at an even higher motor torque. 
We attach the flagellum to a spherical cell body and also observe the first 
buckling transition during locomotion. By changing the size of the cell body, 
we vary the necessary thrust force and thereby obtain a characteristic 
relation between the critical thrust force and motor torque. We present a 
sophisticated analytical model for the buckling transition based on a helical rod which quantitatively reproduces the critical 
force-torque relation. Real values for motor torque, cell body size, and the
geometry of the helical filament suggest that buckling should occur in
single bacterial flagella. We also find that the orientation of 
pulling flagella along the driving torque is not stable and comment on 
the biological relevance for marine bacteria.
\PACS{
      {PACS-key}{discribing text of that key}   \and
      {PACS-key}{discribing text of that key}
     } 
} 
\maketitle
\section{Introduction} \label{intro}

Many bacteria such as \textit{Escherichia coli} and 
\textit{Sal\-mo\-nella typhimurium} swim by rotating a bundle of helical
flagella \cite{Berg2004}. Nature's simple and ingenious solution for 
locomotion at low Reynolds number has already inspired researchers to 
apply rotating flagella to perform such diverse tasks as pumping 
fluid \cite{Darnton2004} or manufacturing nanotubes \cite{Hesse2009}. 
Even artificial helical flagella already exist \cite{Zhang2009}.

The flagellum in the bundle consists of three parts; the rotary motor, a 
short and very flexible proximal hook that couples the motor to the third 
part, the long helical filament \cite{Berg2004,Turner2000,Darnton2007a}. 
The motors are embedded at 
different locations of the cell wall so that the flagella have to bend 
around the cell body to form a bundle. Bacteria such as 
\textit{Escherichia coli} and \textit{Sal\-mo\-nella typhimurium} use this 
bundle to perform a run-and-tumble motion which enables them to follow 
a chemical gradient (chemotaxis). After swimming for about
$1 \second$, the sense of rotation of one motor reverses and the 
attached flagellum leaves the bundle. It goes through a sequence of 
polymorphic conformations until the motor reverses its rotational 
direction again. The flagellum returns to its original or normal 
helical form and rejoins the bundle. During this tumbling event, 
which lasts for about $0.1\second$, the bacterium changes its swimming 
direction randomly. It is interesting that flagella of cells, which are 
clued to a surface, do not bundle\ \cite{Darnton2007a,Turner2000}. 
Furthermore, each flagellum is reported to be relatively rigid and 
visible deformations due to rotation are not reported\ 
\cite{Darnton2007a}.
In general, videos show a complex behavior of a single flagellum when it
interacts with other flagella, with the wall, or when it goes through
different polymorphic conformations \cite{Turner2000}.
In this context, recent articles study the synchronization and bundling 
of two or more flagella due to hydrodynamic interactions\ 
\cite{Kim2003,Kim2004,Reichert2005,Janssen2011}.

The polymorphism of the flagellum is a fascinating and intensively 
studied subject
\cite{asakura1970,calladine1975,Macnab1977,Hotani1982,Hasegawa1982,Goldstein2000,Srigiriraju2005,Darnton2007,Wada2008}.
Using a coarse-grained molecular model, we recently addressed the
question why the normal polymorphic state is realized in a flagellum
\cite{Speier2011}
and also developed a simple model for flagellar growth \cite{Schmitt2011}.
Based on an extended Kirchhoff theory for the helical filament, we modeled 
the polymorphism of the flagellum \cite{Vogel2010} and were able to
reproduce experimental force-extension curves where a polymorphic 
transition is induced by an external force\ \cite{Darnton2007}.

\begin{figure}
\begin{center}
\includegraphics[width=0.7\columnwidth]{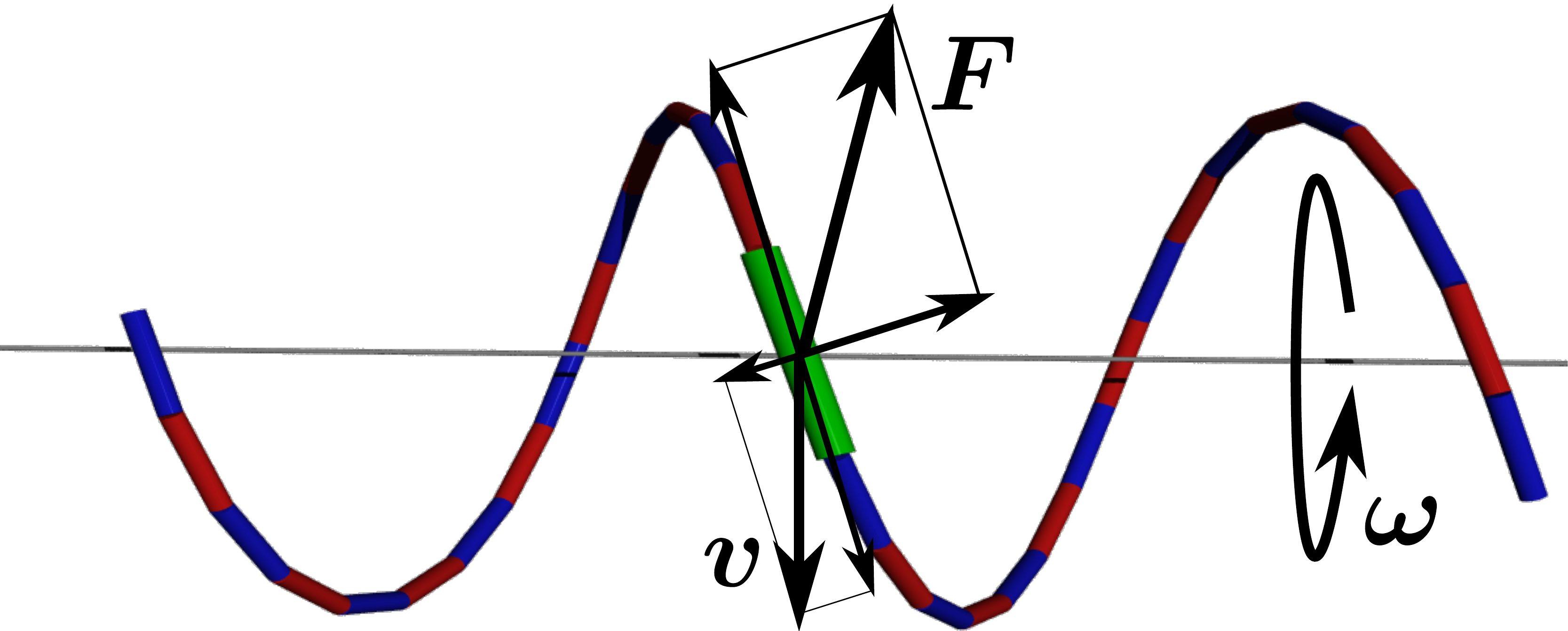}
\end{center}
\caption{Each segment of a rotating helical flagellum experiences
a frictional force $\vec F$ that is not antiparallel to the local 
velocity $\vec v$ due to the anisotropic friction of a rod. Whereas the 
force component perpendicular to the helix axis averages to zero 
over one helical turn, the parallel component adds up to 
the thrust force. For a detailed treatment see appendix\ \ref{app friction}.
}
\label{fig: sketch thrust}
\end{figure}

In this article we concentrate on the normal form of a single bacterial 
flagellum, model it by the discretized version of Kirchhoff's elastic-rod 
theory and develop a coarse-grained approach for driving the helical 
filament by a motor torque. A rotating helical flagellum produces a 
thrust force as explained in fig.\ \ref{fig: sketch thrust} that adds up
along the filament and then pushes the cell body forward. We report two 
buckling instabilities of a fixed helical filament for increasing
motor torque. The first instability occurs in the biologically relevant 
regime. The straight helical filament starts to bend under the influence 
of the acting thrust force similar to a rod that buckles under 
its own load. The buckling instability is visible as a supercritical 
Hopf bifurcation in the thrust force. It also occurs when the filament 
is allowed to move by attaching it to a load particle.
We will develop an analytical model based on a rigid 
helical rod that explains the buckling transition and reproduces 
quantitatively the critical force-torque relation from our simulations.

\begin{figure}
 \begin{center}
    \includegraphics[width=.9\columnwidth]{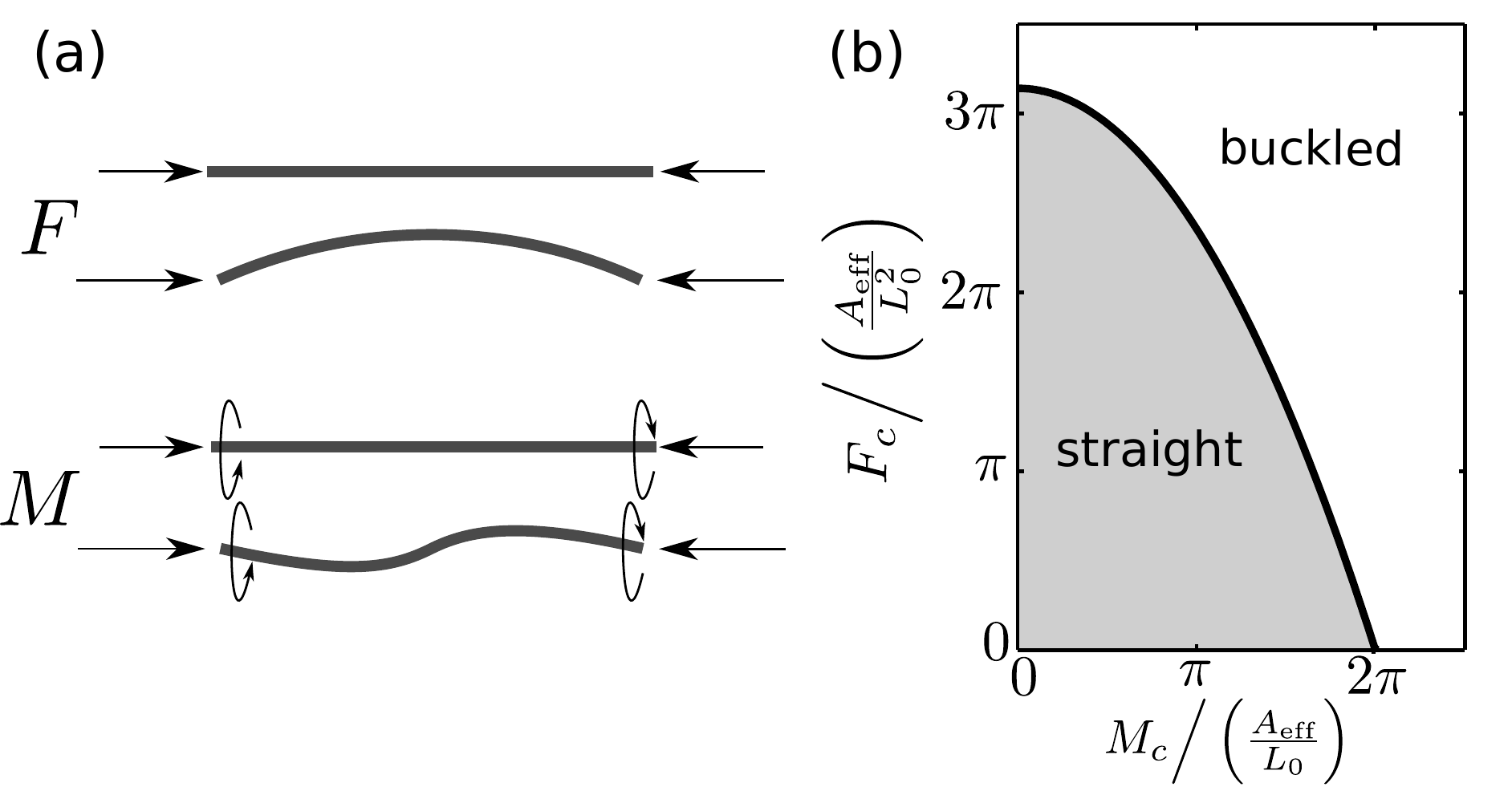}
 \end{center}
 \caption{a) An elastic rod buckles under the influence of a compressional
force $F$ and an external torque $M$. b) The critical values $F_c$ and 
$M_c$ at which buckling occurs obey a characteristic relation. The graph
depicts relation\ \eqref{eq: classical buckling} valid for a rod with 
fixed ends.
}
\label{fig: buckling}
\end{figure}

An elastic rod buckles under the influence of a compressional force and  
an external torque acting at both ends of the rod 
[fig.\ \ref{fig: buckling}(a)]. This is one of the first examples for
a bifurcation and Euler was the first to provide the theory for the 
critical load force at zero torque. In general, critical force 
and torque for a rod with fixed ends obey the relation \cite{Love1944}
\begin{align}
 \pi^2&= F_c \left(\frac{L_0^2}{A}\right) +  \frac{1}{4}M_c^2 
\left(\frac{L_0}{A}\right)^2,
\label{eq: classical buckling}
\end{align}
where $L_0$ is the rod length and $A$ its bending rigidity 
[fig.\ \ref{fig: buckling}(b)]. Note that 
eq.\ (\ref{eq: classical buckling}) does not depend on the compressibility 
or the torsional rigidity of the rod.

Similar buckling or elastic instabilities occur in the dynamics of rods 
at low Reynolds number. Here one typically applies a torque at one end of the 
filament. The rod rotates and the applied torque is balanced by frictional 
forces and torques continuously distributed along the filament. 
Wolgemuth \textit{et.al.} investigated a rod with one clamped and one free 
end rotating around its axis. They observed two regimes separated by a 
supercritical (\textit{i.e.} continues) Hopf bifurcation. When the 
rotational frequency exceeds a critical value, the straight filament
starts to bend and performs a whirling motion \cite{Wolgemuth2000}.
In Brownian dynamics simulations Wada and Netz observed for the same 
conditions a subcritical (\textit{i.e.} discontinuous) Hopf bifurcation 
where the strongly bent filament nearly folds back on itself \cite{Wada2006}. 
On the other hand, a rod tilted with respect to the rotational axis
bends slightly due to friction at low rotational velocity. At a 
critical value, a discontinuous transition to a helical rod shape 
occurs \cite{Manghi2006}.

In this article we treat buckling instabilities for the biologically 
relevant helical filament. The problem is more complex due to the 
characteristic rotation-translation coupling and the fact that we
do not fix the orientation of the helical filament. The content of the
article is organized as follows. In sect.\ \ref{sec:1} we explain how we
model the motor-driven bacterial flagellum and how we perform the
simulations. In sect.\ \ref{sec: The motor-driven helical filament} we
present and discuss our numerical results for both buckling instabilities 
for the fixed filament and then address the first buckling instability
during locomotion of the filament. In sect.\ \ref{sec.buckling theory}
we formulate a buckling theory for a helical rod and show that
it quantitatively reproduces the critical force-torque relation in 
the biologically relevant regime. We close with a summary and conclusions.

\section{Modeling the motor-driven bacterial flagellum} \label{sec:1}

We start with a short review of the elasticity model and the dynamics
of a helical filament in sects.\ \ref{subsec.elast} and \ref{subsec.dynamics} 
following our previous work\ \cite{Vogel2010} and then explain how we model 
the motor-driven hook in sect.\ \ref{subsec.hook}. A summary of the simulation
parameters follows in sect.\ \ref{subsec.simulpar}.

\subsection{Elasticity model of a helical filament}
\label{subsec.elast}

\begin{figure}
\begin{center}
\includegraphics[width=0.7\columnwidth]{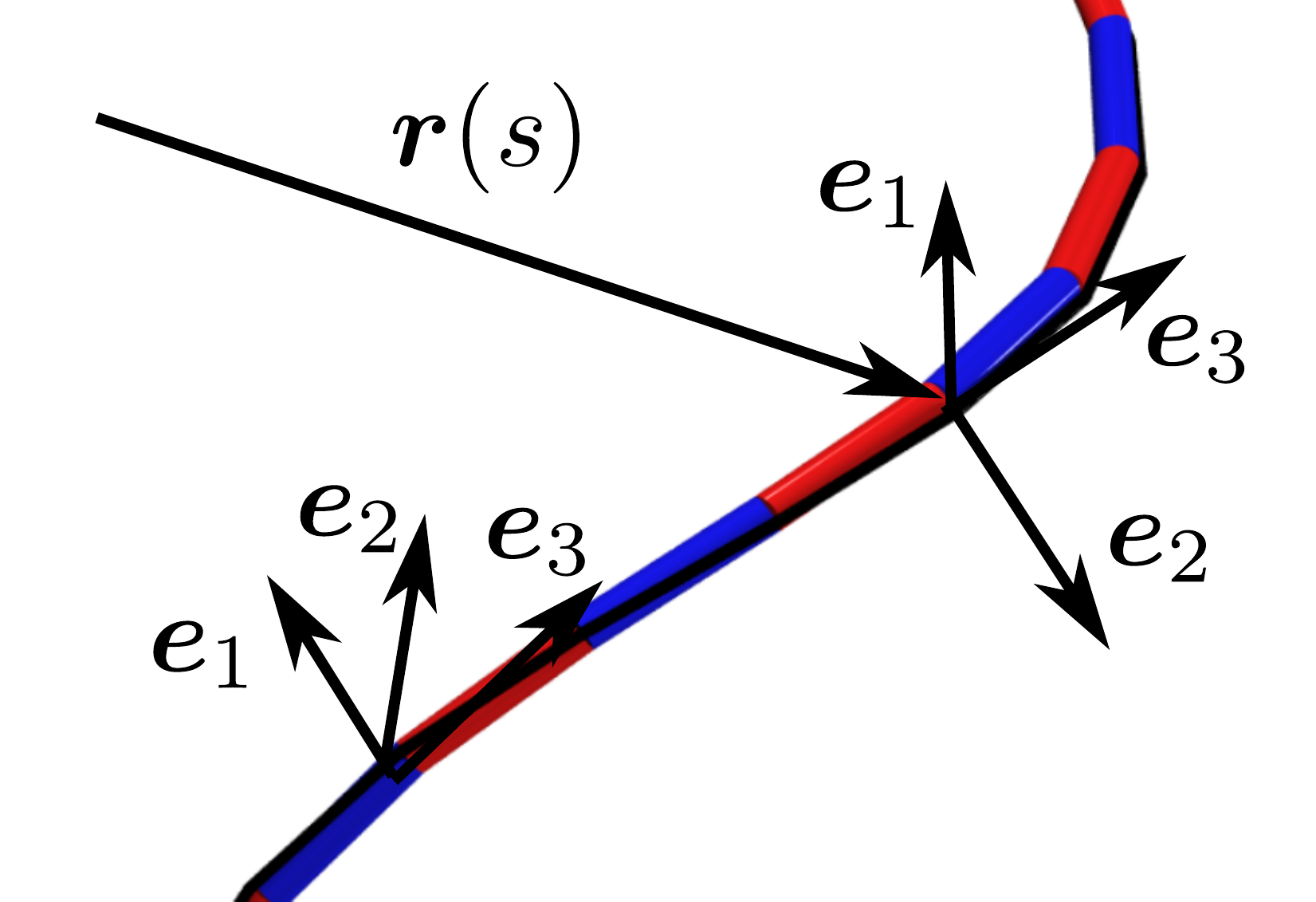}
\end{center}
\caption{The kinematic variables of a slender elastic rod are the 
space curve $\vec r(s)$ of its center line and the material 
frame $\{\vec e_1,\vec e_2,\vec e_3\}$ attached at each point of the 
center line.
}
\label{fig.material}
\end{figure}

We describe the conformation of the helical filament with contour 
length $L$ by the space curve of its center line $\vec r(s)$, where $s$ is the 
arc length. In addition, we attach a material frame of three orthogonal 
unit vectors $\{\vec e_1,\vec e_2,\vec e_3\}$, to each point along the filament so that $\vec e_3$ points along the tangent of $\vec r(s)$
(see fig.\ \ref{fig.material}). The generalized Frenet-Serret
equations transport the material frame along the filament,
\begin{align}
\@_s \vec e_i = \vec \Omega \times \vec e_i ,
\end{align}
where $\partial_s$ means derivative with respect to $s$. We can 
characterize any conformation by the angular 
strain vector $\vec \Omega = (\Omega_1,\Omega_2,\Omega_3)$ given in 
components with respect to the local material frame. Along an ideal 
helical filament,
spontaneous curvature $\kappa_0$ and torsion $\tau_0$ are constant.
As material frame for the equilibrium shape of the helical filament, 
we choose the Frenet frame which consists of the tangent vector 
$\vec t= \vec e_3$, the normal vector 
$\vec e_1 = \vec n = \partial_s \vec t / \kappa_0$, and the binormal 
vector $\vec e_2 = \vec b = \vec t \times \vec n$. The strain vector
then becomes $\vec \Omega=(0,\kappa_0,\tau_0)$. Further parameters
of an ideal helix are the pitch $p=2 \pi \tau_0/(\kappa_0^2+\tau_0^2)$ and 
radius $R=\kappa_0/(\kappa_0^2+\tau_0^2)$. The ratio of pitch and 
circumference, $p/2\pi R = \tan\alpha$, defines the pitch angle $\alpha$.




The total elastic free energy of the filament consists of two
contributions:
\begin{equation} 
\label{Glg_freieEnergie}
{\mathcal F}= \int_0^L (f_{\mathtt{cl}} + f_{\mathtt{st}} )\mathrm d s
\end{equation}
The first term is Kirchhoff's classical theory for bending and 
twisting,
\begin{equation} 
\label{Glg_freieEnergie1}
f_{\mathtt{cl}}= \frac{A}{2} (\Omega_1)^2 + \frac{A}{2} (\Omega_2-\kappa_0)^2 
+ \frac{C}{2} (\Omega_3-\tau_0)^2 ,
\end{equation}
where we introduced the bending rigidity $A$ and the torsional 
rigidity $C$ \cite{Love1944,Landau1986}. Instead of implementing a 
constraint for the inextensibility of the filament in our simulations,
we also add a stretching free energy with line density
\begin{align}
f_{\mathtt{st}} = \frac{K}{2} \left(\@_s \vec r\right)^2 .
\label{Glg_freieEnergie2}
\end{align}
We choose the spring constant $K$ such that the changes in the filament 
length are below 1.5 \%. The filament is inextensible to a good 
approximation.

%



\subsection{Dynamics of the helical filament}
\label{subsec.dynamics}

We mostly performed deterministic simulations, only in a few cases
we have added thermal fluctuations. We formulate Langevin equations 
for the location $\vec r(s)$ and intrinsic twist $\phi(s)$ of the 
helical filament. At low Reynolds number elastic force per unit length, 
$\vec f_\mathtt{el} = - \delta \cal F/\delta \vec r$, and thermal force 
$\vec f_\mathtt{th}$ are balanced by viscous drag. The same applies 
to the elastic torque per unit length, $m_\mathtt{el}=-\delta 
\cal F/\delta \phi$ and thermal torque $m_\mathtt{th}$. Using resistive-force
theory, we introduce local friction coefficients 
$\gamma_\parallel, \gamma_\perp$ and $\gamma_R$ (see Appendix\ 
\ref{app friction}) and arrive at the Langevin equations
\begin{align}
 \left[\gamma_\parallel \vec t \otimes 
\vec t + \gamma_\perp(\mat 1- \vec t \otimes \vec t )\right]
\vec v=&\vec f_\mathtt{el}+ \vec f_\mathtt{th}
\label{eq.dynamic1}\\
\gamma_R \omega = & m_\mathtt{el} + m_\mathtt{th} .
\end{align}
Here $\vec v=\@_t \vec r$ is the translational velocity, 
$\omega = \@_t \phi$ the angular velocity about the local tangent 
vector $\vec t=\vec e_3$, and $\otimes$ means tensorial product. 
The anisotropic friction tensor acting on $\vec v$ in 
Eq.\ (\ref{eq.dynamic1}) couples rotation about the
helical axis to translation and thereby creates the thrust force that
pushes the bacterium forward as illustrated in 
fig.\ \ref{fig: sketch thrust} \cite{Purcell1977}. 
Experiments show reasonable agreement with the approach of 
resistive-force theory \cite{Chattopadhyay2006,Chattopadhyay2009}.
Finally, the thermal force $\vec f_\mathtt{th}$ and torque $m_\mathtt{th}$ are
Gaussian stochastic variables with zero mean,
$\langle \vec f_\mathtt{th} \rangle = \vec 0$ and 
$\langle m_\mathtt{th} \rangle = 0$. Their variances obey the 
fluctuation-dissipation theorem and therefore read
\begin{align}
 \m{\vec f_\mathtt{th}(t,s) \otimes \vec f_\mathtt{th}(t',s')} &= 2 k_B T 
\delta(t-t')\delta(s-s') \\
 & \times \left[\gamma_\parallel \vec t \otimes \vec t + \gamma_\perp
(\mat 1- \vec t \otimes \vec t )\right],
\nonumber \\
\m{m_\mathtt{th}(t,s)m_\mathtt{th}(t',s')}&= 2 k_B T \delta(t-t')
\delta(s-s')\gamma_R,\\
\m{m_\mathtt{th}(t,s)\vec f_\mathtt{th}(t',s')}&=\vec 0.
\end{align}

In our simulations we use a discretized version of the dynamic equations 
following our earlier work \cite{Reichert2006,Vogel2010} (see also 
Ref.\ \cite{Chirico1994,Wada2007a,Wada2007}). We discretize the center line 
$\vec r(s)$ of the filament by introducing $N+1$ beads 
at locations $\vec r^{(i)}=\vec r(s=i\cdot h)$ and with nearest-neighbor 
distance $h$. To every bead we attach the material frame 
$\{\vec e^{(i)}_1,\vec e^{(i)}_2,\vec e^{(i)}_3\}$ ($i=0,\dots,N$) and 
approximate the tangent vector by
\begin{align}
 \vec e_3^{(i)} &=\frac{\vec r^{(i)}-\vec r^{(i-1)}}{|\vec r^{(i)} 
-\vec r^{(i-1)}|}.
\end{align}
The transport of the material frame along the filament occurs in two steps:
First, we rotate about the bond direction $\vec e^{(i)}_3$ by an angle 
$\Omega_3^{(i)} h$ to implement intrinsic twist plus torsion. Thereafter, 
we introduce the curvature of the filament, by rotating the
the bond vector $\vec e_3^{(i)}$ of the material frame about
$\vec e_3^{(i)} \times \vec e_3^{(i+1)}$ by an angle 
$\sqrt{\Omega_1^2+\Omega_2^2} h$
into the consecutive direction $\vec e_3^{(i+1)}$. With this procedure the 
free energy densities $f_{\mathtt{cl}}$ and $f_{\mathtt{st}}$ from
Eqs.\ (\ref{Glg_freieEnergie1}) and (\ref{Glg_freieEnergie2}) are discretized
and the functional derivatives of the total free energy,
$\vec f_\mathtt{el} = - \delta \cal F/\delta \vec r$ and 
$m_\mathtt{el}=-\delta \cal F/\delta \phi$, reduce to conventional 
derivatives with respect to $\vec r^{(i)}$ and $\phi^{(i)}$. In addition,
we approximate the tangent vector in the friction tensor in 
Eq.\ (\ref{eq.dynamic1}) by 
$\vec t^{(i)} = (\vec e_3^{(i)} + \vec e_3^{(i+1)})/|\vec e_3^{(i)} + \vec e_3^{(i+1)}|$.

\subsection{The motor-driven hook} \label{subsec.hook}

\begin{figure}
     \begin{center}
      \includegraphics[width=1.\columnwidth]{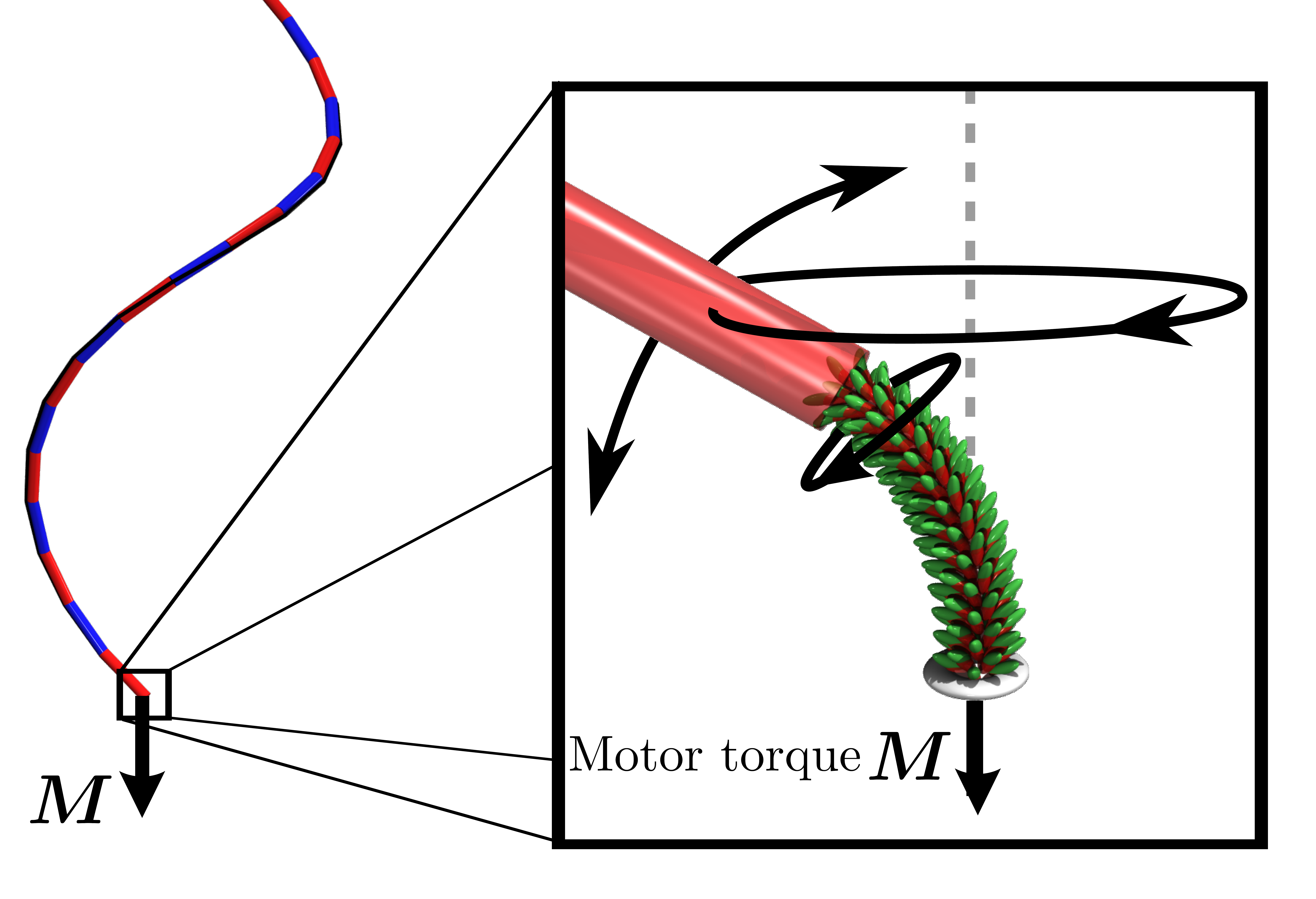}
    \end{center}
\caption{Blow-up: The hook acts as a universal joint between the 
motor embedded in the cell wall and the long helical filament which
retains its full rotational degrees of freedom.
Main picture: The hook is much shorter than the discretization length 
indicated by the blue and red segments of the filament. We do not model
the hook explicitly but let the motor torque act directly on the first 
material frame of the filament which, in principle, can assume any 
orientation in space.
}\label{fig: hook}
\end{figure}

The flagellum is driven by a rotary motor embedded in the cell wall 
of the bacterium. The motor torque is transmitted to the helical filament 
by a short flexible coupling. Because of its shape it is called hook. 
With a well regulated length
of $0.05\micro\meter$ for \textit{E.Coli} or \textit{S.typhimurium} 
and up to  $0.1\micro\meter$ for \textit{R. sphaeroides} it is much shorter 
than the helical filament
\cite{Jones1990,Kobayashi2003,Samatey2004,Shaikh2005}. 
It is also shorter than the discretization length of $h=0.2\micro\meter$
which we can employ in our simulations as indicated in fig. \ref{fig: hook}. 
We, therefore, cannot model the hook in full detail. Instead, we represent
motor and hook by a motor torque that acts directly on one end of the filament 
neglecting the extension of the hook. 

Molecular dynamics simulations showed that the hook bends and twists easily. 
This is possible since conformational changes of molecular bonds require 
only a small amount of energy \cite{Furuta2007}. So the hook itself allows 
the filament to nearly assume any orientation outside the cell. 
Hence, it is comparable to a constant-velocity joint.
The blow-up in fig.\ \ref{fig: hook} illustrates how motor and 
hook act together to drive the filament. The picture also shows the 
rotational degrees of freedom of the filament at the attachment point to 
the hook. The filament can rotate about its local axis, about the axis 
parallel to the motor torque, and towards or away from this axis.

%


The task of the hook is to transmit the motor torque to the filament and
to guarantee its rotational degrees of freedom. In our coarse-grained model,
we implement this task by balancing all the torques acting on the first 
material frame $\{\vec e_1^{(0)}, \vec e_2^{(0)},\vec e_3^{(0)}\}$ that determines
the orientation of one end of the filament:
\begin{align}
\big[\gamma_R h &\vec e_3^{(0)}\vec e_3^{(0)} + \frac{1}{2} \gamma_\perp 
h^3 (\mat 1 - \vec e_3^{(0)} \vec e_3^{(0)}) \big] \vec\omega \\
 = &  
\vec M - A [\Omega_1 \vec e_1^{(0)} + (\Omega_2-\kappa_0) \vec e_2^{(0)} ] 
- C (\Omega_3-\tau_0)\vec e_3^{(0)}.
\nonumber
\end{align}
The material frame rotates with an angular frequency $\vec\omega$. It gives rise
to a frictional torque decomposed into a component along the tangent vector
$\vec e_3^{(0)}$ and perpendicular to it. The length $h$ appears due to the
discretization. The frictional torque is balanced by the motor or external
torque $\vec M = M \vec e_z$, 
which we assume constant throughout the paper, and the elastic 
torque $-\delta {\mathcal F}/ \delta \vec \Omega$.

%

\subsection{Simulation parameters} \label{subsec.simulpar}

For the bending rigidity we use $A=3.5 \pico\newton \micro\meter^2$ 
given in Ref. \cite{Darnton2007} as a typical value for bacterial flagella
and set it equal to the torsional rigidity, $C=A$. Our previous work
showed that this is in agreement with experimental observations\ 
\cite{Vogel2010}.
All other parameters are determined  by the geometry. In our study 
we use the normal state of the bacterial flagellum with spontaneous 
curvature $\kappa_1=1.3\per\micro\meter$ and torsion 
$\tau_1=2.1\per\micro\meter$. 
In the following we study a right-handed helical filament although
the normal state of a real flagellum is left-handed.
We calculate the local friction coefficients
from Lighthill's formulas \cite{Lighthill1976} summarized in 
appendix\ \ref{app friction} as $\gamma_\parallel=1.6\cdot 10^{-3} 
\pico \newton \second \per \micro \meter^2$, $\gamma_\perp=2.8\cdot10^{-3} 
\pico\newton\second\per\micro\meter^2$, and $\gamma_R=1.26\cdot 
10^{-6} \pico\newton\second$, where a filament diameter of about 
$20\nano\meter$ is used. The length of the filament is $L=10\micro\meter$ 
corresponding to approximately four helical turns. The discretization length 
between the beads is chosen as $h=0.2\micro\meter$.

%
%

\section{The motor-driven helical filament}
\label{sec: The motor-driven helical filament}

In this section we study in detail the thrust force that the motor-driven 
helical filament generates both when the actuated end of the filament is 
fixed in space or attached to a larger load particle, which mimics the
cell body. In particular, we describe the buckling transitions by illustrating 
the observed filament configurations. 

It is instructive to shortly look at a completely rigid helical rod first,
which does not exhibit translational motion. In the low Reynolds number 
regime, the angular velocity $\vec \omega$ of the rod and the applied 
torque $\vec M$ obey the linear relation
$ 
\vec M = \mat B   \vec \omega ,
$
where $\mat B$ is the rotational friction tensor. For a long slender 
helix like the normal form of the bacterial filament, one principal axis 
of $\mat B$ points along the helical axis and the eigenvalues in the 
plane perpendicular to this axis are degenerate, in good approximation
\cite{Lighthill1976,Childress1981}. Now there is a formal analogy to the 
motion of the force and torque less spinning top with axial symmetry 
in classical mechanics \cite{Landau1976,Arnold1978}. We just replace the 
constant torque $\vec M$ by the conserved angular momentum and $\mat B$ 
by the moment of inertia tensor. We explain details in appendix 
\ref{app rotating helix}. According to this analogy, the rigid helix in a 
viscous fluids precesses about the constant applied torque while also 
rotating about its helical axis. However, in our simulations we 
observe that as soon as we introduce a finite elasticity of the helical 
rod, the precession is no longer stable and the helical filament 
aligns, for example, parallel to the torque.

\subsection{Force-torque relation and buckling}
\label{sec: Motor driven rotation}

\subsubsection{Discussion of the basic features}
\label{subsec:MDRDiscussion}

\begin{figure*}
\begin{center}
  \includegraphics[width=\textwidth]{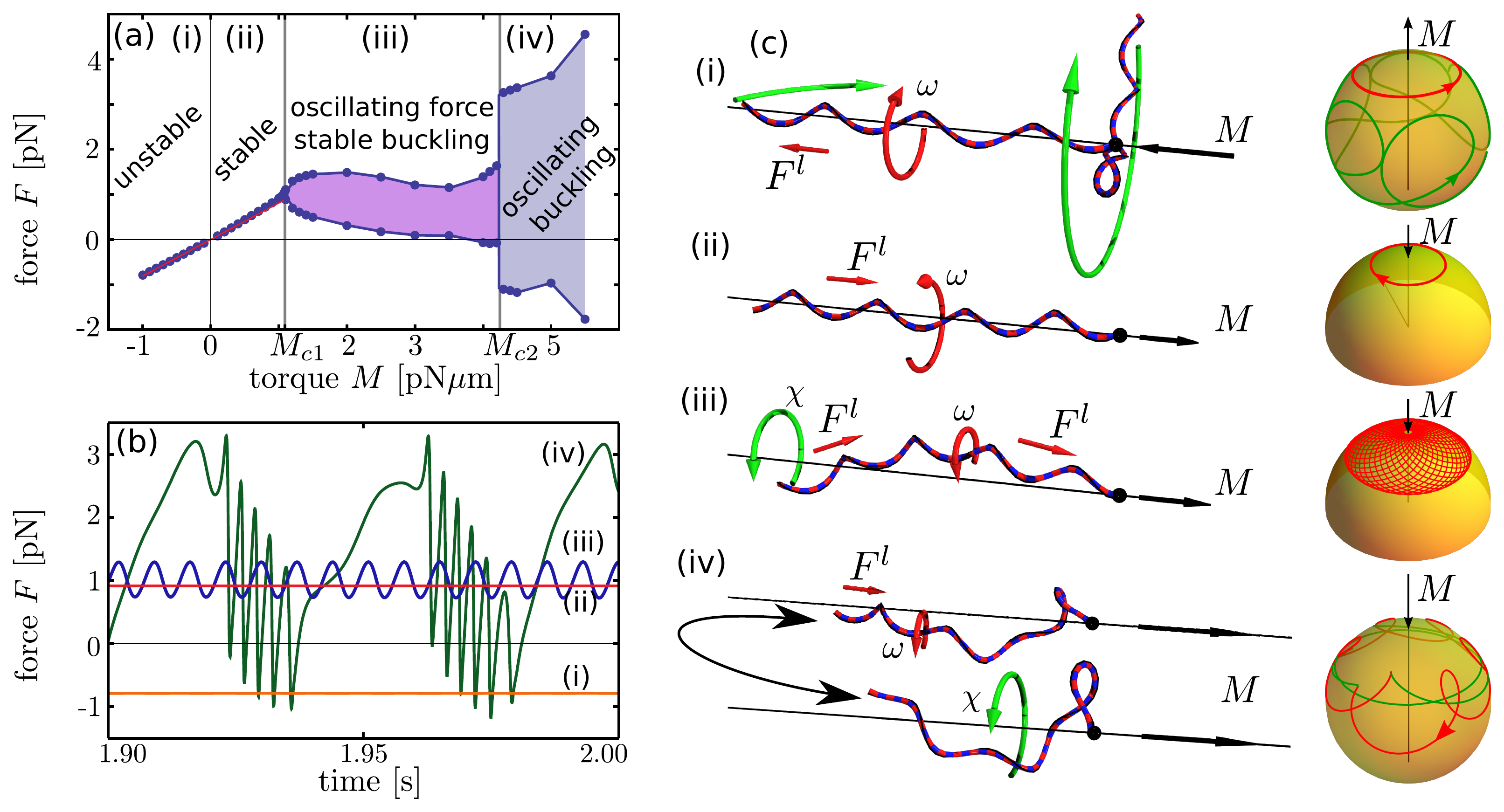}
\end{center}
\caption{(a) Thrust force $F$ versus motor torque $M$. Four different regimes
associated with different configurations of the rotating filament exist. 
In regime (iii) and (iv) the minimum and maximum value of the oscillating
thrust force are shown. A supercritical bifurcation occurs at the critical
torque $M_{c1}\approx1.1\pico\newton\micro\meter$ indicating a buckling
transition. A second bifurcation is visible at 
$M_{c2}\approx 4.2 \pico\newton\micro\meter$. The red line in regimes 
(i) and (ii) follows from resistive force theory.
(b) Thrust force versus time for specific torque values in the four 
different regimes: (i) $M=-1.0\pico\newton\micro\meter$, 
(ii) $M=1.0\pico\newton\micro\meter$, 
(iii) $M=1.2\pico\newton\micro\meter$, and
(iv) $M=4.5\pico\newton\micro\meter$.
(c) Characteristic snapshots of the helical filament in the four regimes.
The red circular arrow and $\omega$ indicate a rotation about the local 
helix axis and $F^{l}$ the local thrust force. The green circular arrow and 
$\chi$ show the precession about the external torque axis. In addition, the
trajectory of the tip of the first tangent vector is indicated:
(i) The green line belongs to the perpendicular orientation of the filament,
(iv) red line: fast rotation about helical axis, green line: slow precession
about motor torque during relaxation of the filament.
}
\label{fig: RotStat}
\end{figure*}

The motor-driven helical filament creates a thrust force. We calculate 
it as the force component on the first bead parallel to the applied 
torque 
$\vec M=M\vec e_z$: $F = -\partial \mathcal F / \partial \vec r^{(0)} 
\cdot \vec  e_z$.
We keep here the bead at a fixed position $\vec r^0=\vec r(0)$ and use the
discretized version of the free energy (\ref{Glg_freieEnergie1}). 
Figure\ \ref{fig: RotStat}(a) plots the resulting thrust force $F$ versus 
the applied torque determined in simulations without thermal noise. We 
discuss the graph in detail.

A positive torque $M$ produces a thrust force that pushes against the
anchoring point of the filament. The thrust force is constant in time
as indicated by the straight line (ii) in fig.\ \ref{fig: RotStat}(b).
The illustration (ii) of fig.\ \ref{fig: RotStat}(c) shows the stable
orientation of the helical filament along the torque $\vec M$. It rotates
about the helical axis with angular frequency $\vec \omega$. The local
thrust force acting along the helix axis is indicated by $F^{l}$. The 
tangent of the filament at the anchoring point is tilted against $\vec M$ 
and the tip moves on a circle, as indicated by the schematic.
Movies for all four types of configurations are available in the 
supplemental material.

A negative torque $M$ generates a negative force that pulls at the anchoring 
point. However, we realized that the orientation of the filament along 
the torque is not stable. For long times the filament turns away from the 
torque axis [green arrow in illustration (i) of fig.\ \ref{fig: RotStat}(c)] 
until it reaches a configuration perpendicular to $\vec M$, where it 
slowly rotates about the local helical axis and slowly precesses about 
$\vec M$. This motion is also visible for the tip of the first tangent vector.
The linear increase of $F$ with $M$ in the regimes (i) and (ii) in 
fig.\ \ref{fig: RotStat}(a) fits well with 
the result from resistive force theory for a perfect helical filament, 
as indicated by the line (see appendix \ref{app friction}). 
Small deviations are visible at higher torques due to elastic
deformations of the helix which enhance the thrust force.

At a critical torque $M_{c1}\approx1.1\pico\newton\micro\meter$ the thrust 
force starts to oscillate as curve (iii) in fig.\ \ref{fig: RotStat}(c)
indicates. Minimum and maximum values of the force are plotted in 
fig.\ \ref{fig: RotStat}(a). They develop continuously from the constant
force at $M_{c1}$ indicating a supercritical Hopf bifurcation. Illustration
(iii) of fig.\ \ref{fig: RotStat}(c) shows a buckled configuration that
rotates about the local helix axis with frequency $\omega$ and precesses
with frequency $\chi$ about the motor torque $\vec M$ keeping its shape
fixed. The trajectory of the tip of the first tangent vector reflects this 
motion.
A straightforward explanation is that the helical filament buckles
under the thrust force generated by the rotating filament. The
force adds up from the free to the fixed end of the filament and puts the
filament under compressional tension. This is similar to a rod that 
buckles under its own gravitational load \cite{Love1944,Landau1986}. 
In sect.\ \ref{sec.buckling theory} we
will develop a theory for this buckling transition which is quite involved.
Finally, at a critical torque value of 
$M_{c2}\approx 4.2 \pico\newton\micro\meter$ a second bifurcation occurs 
in the force-torque relation of 
fig.\ \ref{fig: RotStat}(a). The buckled state itself becomes unstable, 
visible by the fast oscillations of the thrust force in
fig.\ \ref{fig: RotStat}(b). The buckled configuration is compressed
until the fixed end becomes perpendicular to the motor torque. At this
point the fast rotation about the local helical axis stops and the thrust 
force averaged over one fast period is approximately zero. Now the strongly
bent configuration of the filament relaxes slowly and precesses about the
applied torque $\vec M$ [second configuration in 
fig.\ \ref{fig: RotStat}(b)(iv)]. The thrust force on the anchoring point
slowly increases. When the filament is sufficiently relaxed, it starts again
its fast rotations about the local helix axis and the whole cycle repeats.

\subsubsection{Discussion of additional features}\label{sec: buckling - add. feat.}
\begin{figure}
\begin{center}
 \includegraphics[width=\columnwidth]{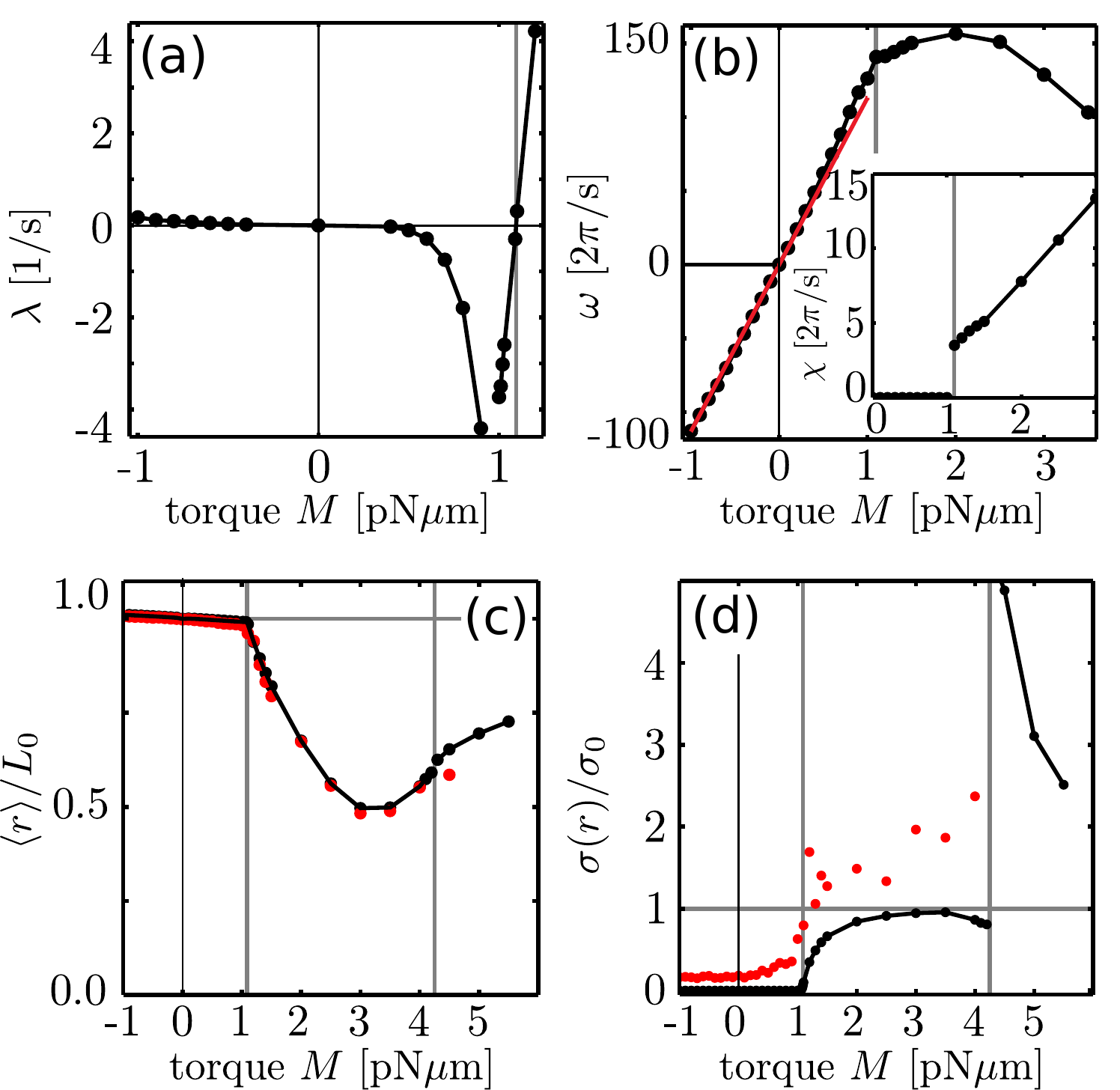}
\end{center}
\caption{(a) Relaxation rate $\lambda$ of the elastic free energy versus
applied torque $M$ for a small disturbance of the aligned state where the
filament is parallel to the torque direction. (b) Angular velocity $\omega$
and precession frequency $\chi$ versus torque $M$. The supercritical 
bifurcation at $M_{c1}$ is clearly visible. The red line is calculated with 
resistive force theory. (c) Mean end-to-end distance $\langle r \rangle$ 
in units of the helix length $L_0$ versus $M$. The red dots are results from
Brownian dynamics simulations with thermal noise included.
(d) Standard deviation $\sigma(r)$ of the end-to-end distance in units of
$\sigma_0=R/\sqrt{2}$ versus $M$. Thermal noise (red dots) leads to 
fluctuations about the mean value.
}
\label{fig: RotStatII}
\end{figure}

The reported supercritical Hopf bifurcation is also visible in
other quantities besides the thrust force. We discuss here additional
properties of the motor-driven helical filament.

To quantify the stability of the filament aligned parallel to the motor 
torque axis, we recorded the temporal evolution of the elastic free energy
starting from a small disturbance of the aligned state and fit it to the form
\begin{align}
 \lvert\mathcal F-\mathcal F^0\rvert &\approx \delta \mathcal F^0 
\exp(\lambda t) \sin( \omega t) .
\end{align}
Here $\omega$ is the angular velocity of the rotating helix leading to
oscillations in $\mathcal F$ and $\lambda$ is the reorientation rate.
The result for $\lambda$ is plotted in fig.\ \ref{fig: RotStatII}(a). 
For positive $M$ 
below the critical torque, the negative $\lambda$ indicates the stable 
aligned state. For small $M$ a reorientation of the filament could not
be detected within the simulation time. Frictional forces are small
and hardly deform the helix which, therefore, just precesses about the
applied torque. Nevertheless, to record the thrust\ force-torque relation,
we always started from an aligned state at $M=1 \pico\newton\micro\meter$
and then changed the driving torque to the desired value and let the elastic 
free energy relax to its stationary value, where we finally recorded the 
thrust force. The small positive $\lambda$ for $M<0$ indicates the slow 
reorientation of the filament towards the perpendicular configuration
The supercritical Hopf bifurcation is located where $\lambda$ changes sign 
from negative to positive.

Figure\ \ref{fig: RotStatII}(b) shows the angular frequency $\omega$ for 
rotations about the local helix axis as a function of $M$. The linear 
regime belongs to
the aligned state, deviations from it occur in the buckled state. The 
precession frequency $\chi$ for rotations of the whole filament about the
torque axis is plotted in the inset. A non-zero $\chi$ corresponds to
the buckled state.

Finally, figs.\ \ref{fig: RotStatII}(c) and (d) plot the mean end-to-end 
distance $\langle r\rangle$ of the helix and its standard deviation $\sigma$
as a function of $M$, respectively. They are defined as
\begin{align}
 \langle r\rangle &= \lim_{T\to\oo} \frac{1}{T} \int_0^T \lvert \vec r (s=L) 
- \vec r(s=0)\rvert \d t\\
\sigma^2 &= \left\langle (r -\langle r\rangle)^2 \right\rangle .
\end{align}
Whereas $\langle r \rangle$ is continuous at both bifurcations, the standard
deviation displays a pronounced discontinuity at the second bifurcation
in agreement with the behavior of the thrust force. We write $\sigma$ in 
units of $\sigma_0 = R/\sqrt{2}$, where $R$ is the helix radius. $\sigma_0$ is 
the maximum value of $\sigma$ in regime (iii) where the buckled helix has a 
constant shape but the free end of the filament rotates on a circle with 
radius $R$. The strong increase of $\sigma$ in regime (iv) is due to the
oscillating buckled state.

The rotating filament also experiences thermal forces due to the viscous
environment. However, since the persistence length 
$A/k_B T \approx 1 \milli\meter$ calculated from the bending rigidity $A$
is much larger than the filament length of $10 \mu\mathrm{m}$, we do expect
that our results are robust against thermal fluctuations. This is 
confirmed by the end-to-end distance $\langle r \rangle$ in fig.\ 
\ref{fig: RotStatII}(c) (red dots) which agrees with the deterministic
simulations. The standard deviation in \ref{fig: RotStatII}(d) indicates
some fluctuations. Below the buckling transition we can directly connect
them to compressional fluctuations using the spring constant of the 
helical filament, $A/(R^2L)$, calculated in our earlier article 
\cite{Vogel2010}.
The equipartition theorem gives 
$\sigma / \sigma_0 \approx 0.15$ in good agreement with the simulated 
value of 0.18. In the buckled state, the helical filament has more
opportunities to fluctuate around which explains the further increase
of $\sigma$.
%
%
%
%
Furthermore, we observe strong fluctuations of the thrust force in our
simulations which result from the delta correlated stochastic forces
acting on the fixed first bead of the discretized filament. An average
over these fluctuations agrees with the deterministic case (data not shown). 
The fluctuations will also be smoothed out in an experiment which performs 
some temporal average during measurement.

\subsection{Buckling instability during locomotion}
\label{subsec.locomotion}


So far we have studied the situation where one end of the filament is 
fixed in space so that it cannot translate. However, rotating flagella
push the cell body of a bacterium forward so that it moves. We mimic 
this scenario by attaching the filament to a bead of radius $a$ which,
for simplicity, can only move along the $z$ direction. The thrust force $F$
generated at the attached end of the filament is then used to push the
sphere forward acting against the Stokes friction force. We observe
similar thrust force-torque relations as for the case of a fixed filament.
The aligned state is again unstable for negative torque and possesses
a larger reorientation rate which might have biological relevance as 
we discuss in sect.\ \ref{sec.cons}.
For positive torque, the aligned state is stable and the thrust force 
grows linearly in the driving torque $M$ until the Hopf bifurcation occurs
at a critical value $M_{c1}$ indicating the buckling instability. 

\begin{figure}
\begin{center}
 \includegraphics[width=.7 \columnwidth]{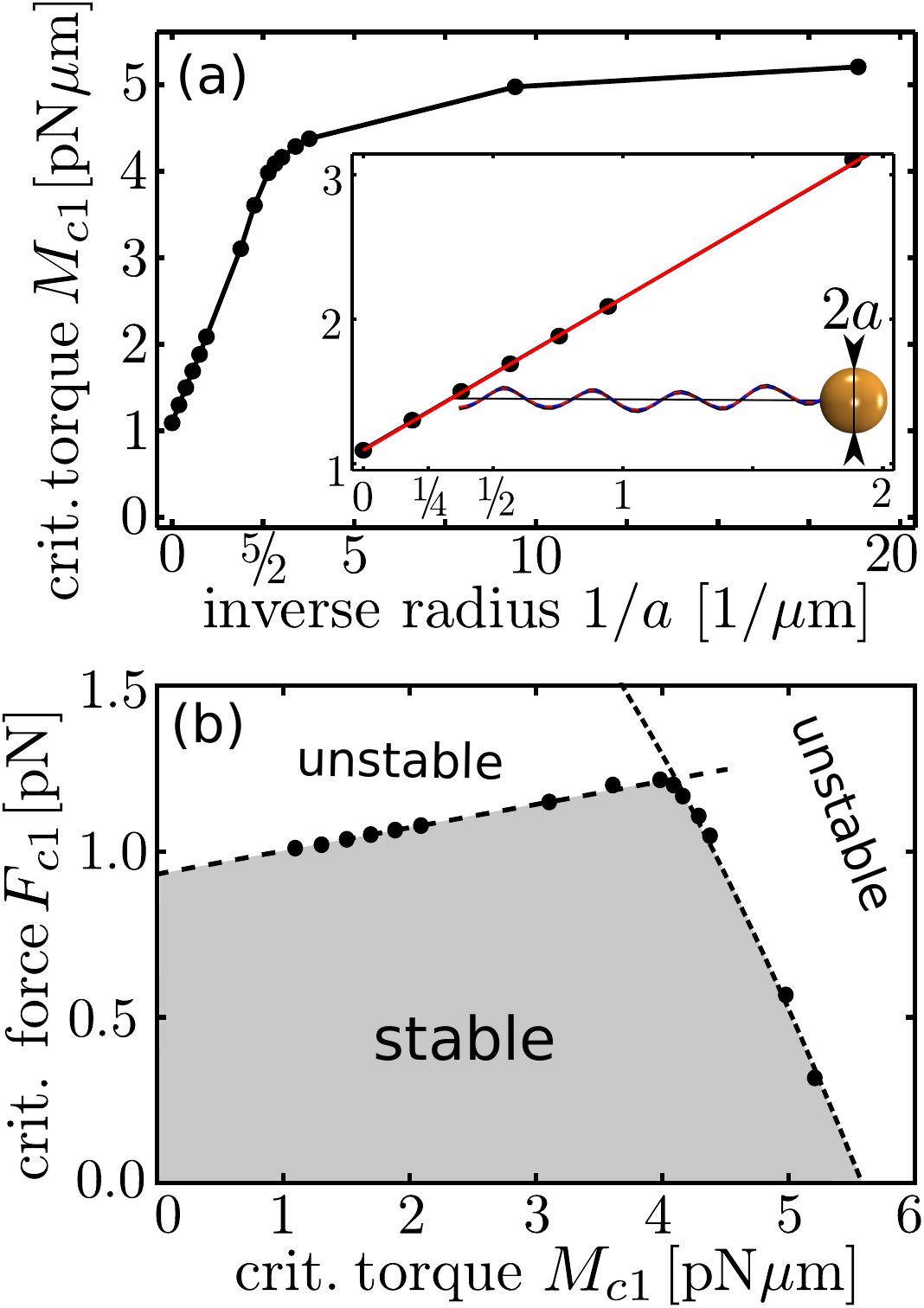} 
\end{center}
\caption{Buckling transition for a helical filament attached to a
bead of radius $a$ that can move along the $z$ direction. (a) Critical
torque $M_{c1}$ as a function of inverse bead radius $1/a$. Inset:
Blow-up for the biologically relevant regime. 
(b) Critical force $F_{c1}$ versus critical torque $M_{c1}$.
}
\label{fig: dynamic}
\end{figure}

Figure\ \ref{fig: dynamic}(a) shows the critical torque $M_{c1}$ as a function 
of the inverse bead radius $1/a$. From $1/a = 0$, which corresponds to
the fixed filament, the critical torque increases linearly in $1/a$ and then
at $a^{-1}\approx 5/2$ turns into a slow growth towards the value for the 
freely swimming helix, i.e., $1/a \rightarrow \infty$. In the biological
relevant case with the cell body size $a\approx0.5\cdots2 \micro\meter$, the
linear dependence of the critical torque on $1/a$ can be derived based on
the fact that the critical thrust force $F_{c1}$ is nearly constant, as we 
show in fig.\ \ref{fig: dynamic}(b). So the velocity $v=F_{c1}/(6\pi \eta a)$ 
is so slow that the buckling transition is hardly influenced by 
the motion of the helical filament with the attached bead. Now, force and 
torque on the helix depend linearly on velocity $v$ and angular velocity 
$\omega$ (see appendix \ref{app friction}). Eliminating $\omega$ and setting 
$v=F_{c1}/(6\pi \eta a)$ at the buckling transition, one arrives at
\begin{equation}
 M_{c1} = -\frac{B_\parallel}{C_\parallel} F_{c1} + 
\left(C_\parallel - \frac{A_\parallel B_\parallel}{C_\parallel}\right)
\frac{F_{c1}}{6\pi \eta} \frac{1}{a}.
\label{eq: M =... mu}
\end{equation}
Here $A_\parallel$, $B_\parallel$, and $C_\parallel$ are the translational, 
the rotational, and coupling friction coefficients parallel to the helical 
axis, respectively. This formula with the coefficients calculated by
resistive force theory (see appendix \ref{app friction}) reproduces the 
linear increase for small $1/a$, as demonstrated by the red line in the 
inset of fig.\ \ref{fig: dynamic}(a).

In fig. \ref{fig: dynamic}(b) we plot the critical thrust force versus
the critical torque. For biologically relevant values $M_{c1}$ between 
$1$ and $2 \pico\newton\micro\meter$ the critical force is indeed nearly 
constant. It only shows a very slow linear increase since frictional 
forces due to the motion of the helix stabilize it against buckling.
At $M_{c1} \approx 4 \pico\newton\micro\meter$ the behavior changes
dramatically. The critical thrust force goes to zero proportional to
$M_{c1}^2$ (see dotted line) following the behavior of a rod that buckles 
under an applied force and torque as described in the introduction. 
In this regime the supercritical Hopf bifurcation becomes subcritical
and hysteresis occurs.
So whereas for small torques buckling is hindered by locomotion, for large
torque the typical quadratic dependence $F_{c1} \propto M_{c1}^2$ is
observed. In the following section, we develop a theory to describe
the observed buckling transition.

\section{Buckling theory for a helical rod} \label{sec.buckling theory}

The goal in this section is to formulate a theory that reproduces 
the force-torque relation in fig.\ \ref{fig: dynamic}(b) for the 
first buckling transition of the helical filament as obtained by our 
simulations. Clearly, this relation cannot directly be explained by the 
theory of a thin elastic rod that buckles under the influence of an external 
force and torque which we shortly mentioned in the introduction in 
Eq.\ (\ref{eq: classical buckling}). 
There are several reasons for this. First, the 
helical filament is not just a simple elastic rod. Second, the external 
force that puts the helix under tension is generated locally by the 
rotation-translation coupling of the helix and accumulated along the 
filament similar to a rod that buckles under its own gravitational
weight. Third, the whole filament moves with a constant velocity which
leads to additional frictional forces and it also precesses about the
external torque in the buckled state. In the following we formulate 
a model based on the theory of a thin elastic helical rod, derive
from it a force-torque relation for the buckling transition, and compare 
it to fig.\ \ref{fig: dynamic}(b).

\subsection{Model equations}

\begin{figure}
\begin{center}
  \includegraphics[width=.7\columnwidth]{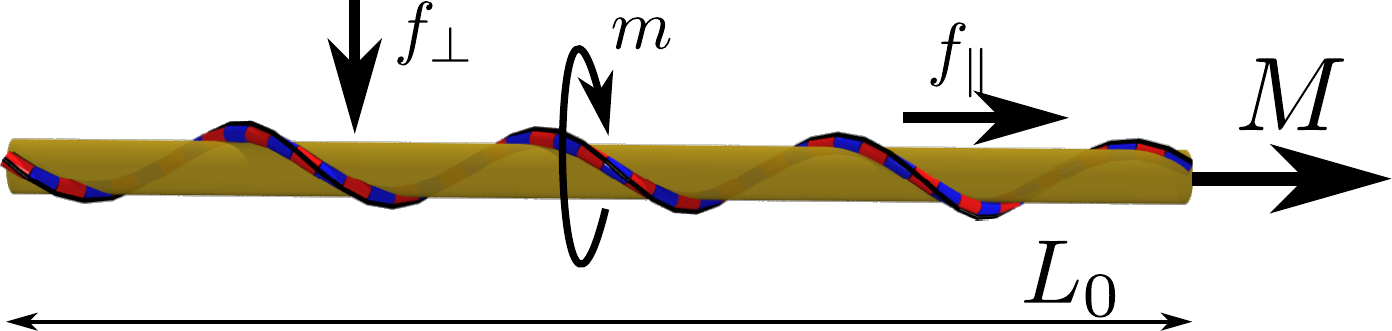}
\end{center}
\caption{The helical filament is approximated by a thin helical rod 
of length $L_0$ that is characterized by the effective bending rigidity $A_\text{eff}$
and local friction coefficients of the helical filament. The applied torque 
$M \vec e_z$ generates the frictional torque $m$ and forces $f_\parallel$ 
and $f_\perp$.
}
\end{figure}

To set up our model equations, we approximate the helical filament by
a thin helical rod where the helicity comes in through the 
rotation-translation coupling in the friction matrix. The length of the
rod, $L_0 = \sin \alpha L$, agrees with the height of the helix,
where $\alpha$ is the pitch angle.
In engineering science the buckling of helical springs is a well known 
problem. If the height of the spring is larger than its radius, one 
approximates the spring by a soft rod with effective bending, shear, and 
compressional rigidity \cite{Haringx1942,Haringx1948,Biezeno1939}.
In our case, in contrast to classical helical spring theory, the pitch of 
the helix is much larger than its radius. We therefore had to generalize 
the theory of helical springs in Ref. \cite{Biezeno1939} to
derive an effective bending rigidity of the helical rod in terms
of the bending and torsional rigidity of the filament:
\begin{align}
\frac{1}{A_\text{eff}}&= \frac{1}{2}\frac{1}{A}\frac{1}{\sin \alpha}
\left(1 + \sin^2 \alpha + \frac{A}{C} \cos^2\alpha\right) 
\label{eq: effbending rigidity}
\end{align}
Details of the derivation are given in appendix \ref{app A_eff}.
%

To address buckling of the helical rod, we start with the balance 
equations for force and momentum acting on a thin elastic rod 
\cite{Love1944,Landau1986} and neglect any inertial contribution
in the low Reynolds number regime: 
\begin{subequations}
\begin{align}
 \vec F' + \vec f &=0,
\label{eq: force balanceI}\\
\vec M' + \vec e_3 \times \vec F + \vec m &=0 ,
\label{eq: force balanceII}
\end{align}
\label{eq: force balance}
\end{subequations}
where $'$ means derivative with respect to the arc length $s$ and
$\vec e_3$ is the local tangent vector.
Here $\vec F$ and $\vec M$ are internal elastic forces and torques acting
along the rod, whereas $\vec f$ and $\vec m$ denote, respectively, external 
force and torque densities due to the applied motor torque and friction
with the surrounding fluid. In addition, boundary conditions are necessary.
At the free end of the rod ($s=L_0$) no external force and torque act, 
so elastic force and torque have to vanish. The end attached to
the sphere can only move in $z$ direction with velocity $v$. 
The external torque $M \vec e_z$ is balanced by the elastic torque 
$\vec M(0)$ and the thrust force on the sphere $F=6 \pi \eta a v$ equals 
the elastic force at the leading end of the rod, $F_z(0)$:
\begin{subequations}
\begin{align}
F_z(0) & = F =6 \pi \eta a v \vec e_z, & \vec M(0) &= M \vec e_z, 
\label{eq: bnd.cnd. 0}\\
\vec F(L_0) &=\vec 0, &\vec M(L_0)&=\vec 0.
\end{align}
\label{eq: force balance bnd. cnd.}
\end{subequations}

After setting up the problem, we have to explain how the different forces 
and torques entering Eqs.\ (\ref{eq: force balance}) look like for the
helical rod close to the buckling transition. The elastic torque $\vec M$ 
is proportional to the angular strain vector $\vec {\Omega}$ written 
in components with respect to the local material frame 
$\{\vec e_1,\vec e_2,\vec e_3\}$: 
\begin{align}
 \vec M & = A_\text{eff} \Omega_1 \vec e_1 + A_\text{eff} \Omega_2 \vec e_2 
+ C_\text{eff} \Omega_3 \vec e_3 , \label{eq: torqu geometry cpl.}
\end{align}
where $A_\text{eff}$ is the effective bending rigidity of 
Eq.\ \eqref{eq: effbending rigidity}. Since buckling theory considers
local displacements of the rod only, the torsional term and the actual value 
of the effective torsional rigidity $C_\text{eff}$ are not important. 
The formulation for $\vec M$ is in full analogy to our presentation in 
sect.\ \ref{subsec.elast}, only the spontaneous curvature and torsion are 
zero for the helical rod which serves as an effective representation 
of the helical filament. In setting up linearized equations in the 
vicinity of the buckling transition, the elastic force $\vec F$ is only 
needed for the unbuckled straight rod oriented along $\vec e_z$. 
Since the external force density $\vec f$ is
constant for the straight rod, as we argue in the next paragraph, 
Eq.\ (\ref{eq: force balanceI}) and boundary conditions
(\ref{eq: force balance bnd. cnd.}) give the linear force profile
\begin{equation}
\vec F(z) = f_{\|} (L_0-z) \vec e_z \enspace \mathrm{with} \enspace
f_{\|} = F/L_0,
\label{eq.linearforce}
\end{equation}
where we introduce $f_{\|}$ as thrust force $F$ divided by the rod length 
$L_0$. We will use it as one parameter in the following.

The straight filament moves with a constant velocity $v \vec e_z$ and
rotates with a constant angular velocity $\omega \vec e_z$. They result,
respectively, in a constant frictional force density $f_{\|} \vec e_z$ and 
a torque density $m \vec e_z$ with
\begin{subequations}
\begin{align}
 f_\parallel & = a_\parallel v + c_\parallel \omega,
\label{eq.frictiontorqueI} \\
           m & = c_\parallel v + b_\parallel \omega,
\label{eq.frictiontorqueII}
\end{align}
\label{eq.frictiontorque}
\end{subequations}
where the frictional coefficient $c_\parallel$ couples translation to 
rotation. Appendix \ref{app friction} gives the coefficients $a_\parallel$, $b_\parallel$, 
and $c_\parallel$ for the helical rod in terms of the parameters of the 
helical filament.
In Eq.\ (\ref{eq.linearforce}) we have already linked $f_{\|}$ to the 
thrust force $F$. From Eq.\ (\ref{eq: force balanceII}) and boundary conditions
(\ref{eq: force balance bnd. cnd.}), we also deduce a linear torque profile
\begin{equation}
\vec M(z) = m (L_0-z) \vec e_z \enspace \mathrm{with} \enspace
m = M/L_0,
\label{eq.lineartorque}
\end{equation}
where we relate $m$ to the applied motor torque $M$ divided by the rod length 
$L_0$. So, $m$ is the second parameter in our problem. 

The buckled rod after the first buckling transition in our simulations
has a constant shape. It rotates about the local tangent vector with
angular velocity $\omega \vec e_3$ and precesses with angular velocity
$\chi$ about the axis of the applied torque leading to a local 
velocity $\chi \vec e_z \times \vec r$. Furthermore, the filament translates
with velocity $v$ along the $z$-direction and the total local velocity
amounts to  $\vec v =  v \vec e_z + \chi \vec e_z \times \vec r$.
In the vicinity of the buckling transition, deformations are small and
in leading order we can identify $v$ and $\omega$ with the values of the
straight rod. Then, the frictional torque along the local tangent vector
is
\begin{equation}
\vec m  = m \vec e_3,
\end{equation}
where $m$ is already given in Eq.\ (\ref{eq.frictiontorqueII}). So, close 
to the buckling instability we can identify $m$ with the applied motor 
torque as in Eq.\ (\ref{eq.lineartorque}). The frictional force density
becomes 
\begin{equation}
 \vec f = f_\parallel \mat P_\parallel\vec e_z + f_\perp \mat P_\perp 
\vec e_z + a_{\perp} \chi \mat P_\perp (\vec e_z \times \vec r),
\end{equation}
where we use the projectors
\begin{align}
  \mat P_\parallel &= \vec e_3 \otimes \vec e_3 \quad \text{and}&
  \mat P_\perp &= \mat 1 - \vec e_3 \otimes \vec e_3
\end{align}
on the directions parallel and perpendicular to the tangent vector $\vec e_3$.
The force density $f_{\|}$ has already been given in 
Eq.\ (\ref{eq.frictiontorqueI}) and
\begin{equation}
 f_\perp = a_\perp v 
\label{eq.frictiontorqueIII}
\end{equation}
characterizes the frictional force density generated perpendicular to the 
local rod axis when the rod moves with velocity $v$. Since the frictional 
coefficient $a_\perp$ is larger than $a_\|$, $f_\perp$ acts against buckling.
Finally, $a_{\perp} \chi$ is the friction due to the precession of the rod.
We note that a term $\mat P_\parallel\cdot (\vec e_z \times \vec r)$ does not 
appear since it does not contribute in leading order to $f_\|$. We also did 
not include the rotation-translation coupling perpendicular to
$\vec e_3$ since the two terms cancel each other in the equations, we
formulate in the following.

We will analyze the buckling transition by first considering the four 
parameters $m,f_\parallel,f_\perp$, and $\chi$ as independent and then 
apply our results to reproduce the force-torque relation of the helical 
rod. 
Buckling occurs when the straight solution $\vec r(z) = (0,0,z)$ of 
Eqs.\ (\ref{eq: force balance}) becomes unstable and a new non-trivial
solution occurs at a certain parameter set. We, therefore, use the 
ansatz $\vec r(z) = (X(z),Y(z),z)$ and seek two equations linear in 
$X$, $Y$, and its derivatives. We start by taking the derivative of 
Eq.\ (\ref{eq: force balanceII}) and use $\vec F' = - \vec f$ to arrive at
\begin{equation}
\vec M'' + \vec e_3' \times \vec F - \vec e_3 \times \vec f + \vec m = 
\vec 0,
\end{equation}
where we insert the concrete formulas for $\vec M$, $\vec F$, $\vec f$,
$\vec m$. We linearize these resulting equations using the identities 
$\Omega_1 \approx -Y''$, $\Omega_2\approx X''$, $\mat P_\parallel\cdot \vec
e_z=\vec e_3\approx(X',Y',1)$, $\mat P_\perp \cdot \vec e_z \approx -(X',Y',0)$, and 
$\mat P_\perp \cdot (\vec e_z \times \vec r)\approx (-  Y, X,0)$, and 
ultimately arrive at
\begin{subequations}
\label{eq: lin buckling}
\begin{align}
0 =&-Y''''+ \partial_z(\hat m (1- \hat z) X'') \nonumber \\
&\qquad- \hat f_\parallel(1- \hat z)Y'' - \hat f_\perp Y' +\hat\chi X, \\
0=&
X''''+ \partial_z(\hat m (1-\hat z) Y'') \nonumber \\
&\qquad+ \hat f_\parallel(1-\hat z)X'' +\hat f_\perp  X' +\hat\chi Y.
  \end{align}
\end{subequations}
Here we introduced the rescaled coordinate $\hat z= z/L_0$ and the 
dimensionless
parameters $\hat m = m L_0^2/A_\text{eff}$, $\hat f_\parallel = 
f_\parallel L_0^3/A_\text{eff}$, $\hat f_\perp=f_\perp L_0^3/A_\text{eff}$, 
and $\hat \chi = \chi a_\perp L_0^4 / A_\text{eff}$. 

Equations \eqref{eq: lin buckling} are quite general and several related 
problems follow from them. 
When forces $\hat f_\parallel$ and $\hat f_\perp$ vanish,
they describe the writhing instability of rotating rods \cite{Wolgemuth2000}. 
For zero torque and precession frequency, and 
$\hat f_\parallel =-\hat f_\perp = \hat f_z$, one arrives at the classical
example of a column that buckles under its own weight 
\cite{Love1944,Landau1986}. A similar problem occurs for microtubuli
that buckle under the action of molecular motors \cite{Karpeev2007}. 
In our case, the force density $f_\|$ that causes buckling points 
along the rod axis and $f_\perp$ stabilizes the straight rod for 
non-zero $v$. In comparison, the column under gravity always experiences a 
force density along the vertical which gives a force component perpendicular
to the rod as soon as it buckles and thereby supports buckling.

We complete the linearized dynamic equations \eqref{eq: lin buckling}
by writing the boundary conditions\ (\ref{eq: force balance bnd. cnd.})
in linearized and reduced form:
\begin{subequations} \label{eq: boundary conditions}
\begin{align}
\label{eq: boundary conditions1}
 X(0)&= 0 & Y(0)&=0,\\
\label{eq: boundary conditions2}
  X''(0)&= - \hat m Y'(0) &Y''(0)&= \hat m X'(0),\\
\label{eq: boundary conditions3}
X''(1) &= 0&Y''(1) &=0,\\
\label{eq: boundary conditions4}
X'''(1)&= 0 & Y'''(1)&=0.
\end{align}
\end{subequations}
The first line means that the attached end of the rod can only move in
$z$ direction and not along the $x$ and $y$ axis.
The second line means that a torque does not act perpendicular to the 
$z$-axis. So if the rod starts to buckle, the local torque $m \vec e_3$ 
has to be equilibrated by a bending moment.
The free end of the rod is torque less and, therefore, the rod does not
bend, as expressed by the third line. Finally, the free end is also force
free and the fourth line follows from Eq. \eqref{eq: force balanceII} 
by setting $\vec F= \vec 0$.

To search for nontrivial solutions of Eqs.\ \eqref{eq: lin buckling} in our
parameter space and thereby identify the buckling transition, we proceeded
as follows. In addition, to the boundary 
conditions\ (\ref{eq: boundary conditions1}) and
(\ref{eq: boundary conditions2}), nontrivial solutions of the buckling
equations\ \eqref{eq: lin buckling} can be characterized by 
$X'(0)$, $Y'(0)$, $X'''(0)$ and $Y'''(0)$. The principal idea is to use 
them to generate solutions of Eqs.\ \eqref{eq: lin buckling} and to 
fulfill the boundary conditions\ (\ref{eq: boundary conditions3}) and
(\ref{eq: boundary conditions4}) at the free end by varying them. 
However, since $X'(0)$ and $Y'(0)$ just 
determine the amplitude of a bent configuration and merely fix the 
rotational degree of freedom about the $z$ axis, they can be chosen 
arbitrary. Instead, we vary two of our four parameters, 
$\hat f_\perp$ and $\hat \chi$, to fulfill the four boundary conditions
at $\hat z = 1$. As a result, for given $\hat m$ and $\hat f_\|$, we 
determine parameters $\hat f_\perp$ and $\hat \chi$ for which non-trivial
solutions of the buckling equations exist and thereby identify
the manifold of bifurcation points in our four-parameter space.
We discuss it in the following section.

\subsection{Discussion}\label{subsec. buckling.discussion}

\begin{figure*}
\begin{center}
 \includegraphics[width=0.8\textwidth]{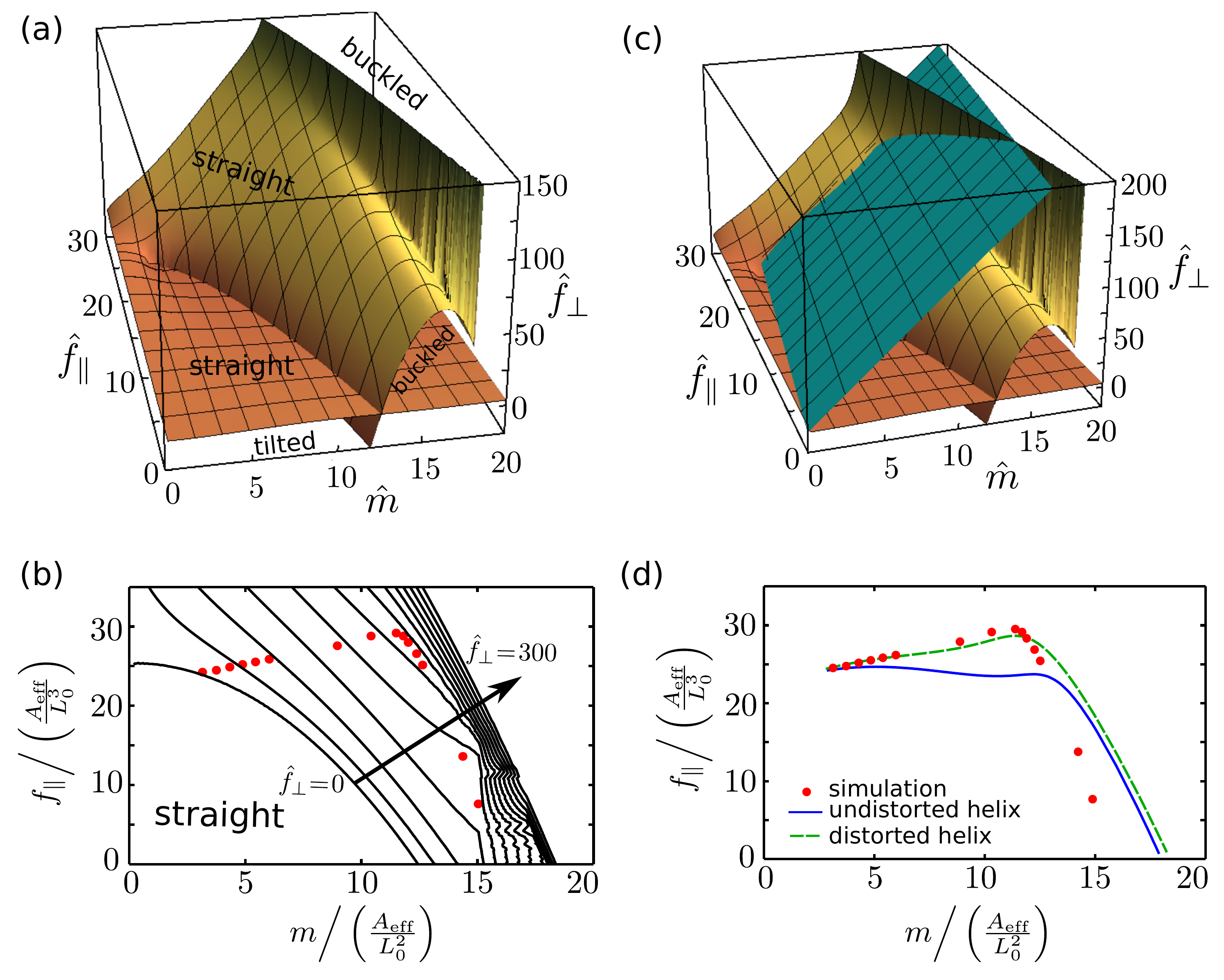}
\end{center}
\caption{(a) Manifold of bifurcation points in the parameter space
$(\hat m, \hat f_\parallel,\hat f_\perp)$. To each parameter triple
belongs a specific value of the precession frequency $\hat \chi$.
(b) Buckling curves $f_\parallel(m)$ for different values of
the perpendicular force ranging from $\hat f_\perp=0$ in steps of 25
to 300. The red dots are the critical forces and torques from
fig.\ \ref{fig: dynamic}(b). (c) The plane pictures relation\
(\ref{eq:plane}) between
$f_\parallel$, $f_\perp$, and $m$ for the helical rod with constant
friction coefficients. Intersecting it with the manifold of bifurcation
points gives the buckling curves in (d). Full blue line: for constant
friction coefficients of the undistorted helix,
dashed green line: torque-dependent friction coefficients.
}
 \label{fig: buckling theory}
\end{figure*}

Figure\ \ref{fig: buckling theory}(a) plots the manifold of bifurcation 
points. To each parameter triple $(\hat m, \hat f_\parallel,\hat f_\perp)$ 
belongs a specific value of the precession frequency $\hat \chi$ which 
we do not discuss further here. At positive $\hat f_\perp$ and for 
small $\hat m$ and $\hat f_\parallel$ the straight configuration of the 
helical rod is stable. 
If we change the sign of 
$\hat f_\perp$, a bifurcation occurs which we 
interpret as an instability of the straight rod when it reorients towards the perpendicular 
configuration. We saw this instability in our simulations when we reversed 
the driving torque as discussed in sects.\ \ref{subsec:MDRDiscussion}
and \ref{subsec.locomotion}.
Here we keep the direction of the torque but reverse the sign of the 
velocity $v$ and thereby the sign of $\hat f_\perp$ in 
Eq.\ (\ref{eq.frictiontorqueIII}) by reversing the chirality of the rod.
The main result is the surface in dark yellow that belongs to the first 
buckling transition observed in our simulations, so at large $\hat m$ the 
rod is buckled. Finally, at $\hat f_\perp \approx 0$ and large $\hat m$ 
a transition between two different configurations of the buckled rod occurs. 
An interesting feature is the ridge in the bifurcation surface. However,
we could not determine any dramatic changes in the buckling of the helical 
rod close to this ridge.

Figure\ \ref{fig: buckling theory}(b) shows buckling curves 
$\hat f_\parallel(\hat m)$ for different values of the perpendicular 
force ranging from $\hat f_\perp=0$ in steps of 25 to 300.
At $\hat f_\perp=0$ the typical parabolic curve of 
Eq.\ (\ref{eq: classical buckling}) occurs. 
At constant but small value of $m$, the critical force $f_\parallel$ 
increases strongly with increasing $\hat f_\perp$. Likewise, one needs large 
forces $\hat f_\perp$ to stabilize the straight helical rod at high torques.
The red dots are the critical forces and torques from 
fig.\ \ref{fig: dynamic}(b) determined in our simulations. We plot them
in reduced units where we calculate $A_\text{eff}$ from 
Eq.\ (\ref{eq: effbending rigidity}).
Note that the buckling curves develop a shoulder at $\hat m$ around
15 for increasing $\hat f_\perp$ due to the ridge in the manifold
of bifurcation points in fig.\ \ref{fig: buckling theory}(a). The two 
simulation points at large $m$ are close to this ridge. We speculate 
that the transition from a supercritical bifurcation observed in our
simulations at low $m$ to a subcritical bifurcation at large $m$ is
connected to the existence of this ridge.

In the rotating helical filament or helical rod, the forces 
$f_\parallel$ and $f_\perp$ and the torque $m$ are related to each other
by Eqs.\ (\ref{eq.frictiontorque}) and (\ref{eq.frictiontorqueIII}). 
Eliminating velocity $v$ and angular frequency $\omega$, one arrives at
\begin{align}
f_\perp &= \frac{a_{\perp}}{a_\parallel b_\parallel-c_\parallel^2} \left(b_\parallel f_\parallel-c_\parallel m\right).
\label{eq:plane}
\end{align}
We give the friction coefficients in terms of the helical parameters
in appendix\ \ref{app friction}. Relation\ (\ref{eq:plane}) defines a plane 
in the parameter space $(\hat m, \hat f_\parallel,\hat f_\perp)$ which 
we intersect in fig.\ \ref{fig: buckling theory}(c)
with the manifold of bifurcation points. The resulting bifurcation curve
is then plotted in fig.\ \ref{fig: buckling theory}(d) as full blue line. 
At low $m$ we have a remarkable quantitative agreement with our 
simulations (red dots) but we miss the slight increase of the critical 
force $f_{\parallel}$. We already discussed that the location of the 
buckling transition is sensitive to small variations in the parameters. 
We also mentioned in the discussion of the simulation results that 
close to the buckling transition the helical filament is slightly deformed.
In Ref.\ \cite{Vogel2010} we calculated the effective spring constant
$A/(R^2 L)$ for the helical filament. It gives a relative compression
of the filament of about $2\%$ for critical forces of $1 \pico\newton$
at the buckling transition, which is negligible.
On the other hand, we apply a torque along the helical axis. As a result,
one end of the helix twists against the other end by an
angle $\Delta \phi/L = M/A$, where we set $A=C$ \cite{Love1944}. One end of
the helical filament is free, so the twist increases linearly from
the free end to the other and the average value is $M/(2A)$.
Due to the twisting, the radius of the helix changes. One can show that
for the average twist angle $M/(2A)$ the inverse radius $R^{-1}$ changes
to $R^{-1} + (\cos \alpha)^{-1} M/(2A)$. Here we keep the pitch angle
$\alpha$ constant, which is confirmed by our simulations. The helical
radius directly influences the friction coefficients $b_{\|}$ and $c_{\|}$
in Eq.\ (\ref{eq:plane}) [see Eqs.\ (\ref{eq: frict coeffi}) and (\ref{eq: frict coeffi BC}) in appendix
\ref{app friction}] and Eq.\ (\ref{eq:plane}) becomes a nonlinear function
in $m = M/L_0$.
Intersecting it with the bifurcation manifold gives the green curve
in fig.\ \ref{fig: buckling theory}(d) which nicely reproduces the
critical force-torque relation for $\hat{m} < 10$.
Our theory also gives the strong decrease of the critical force
$f_{\|}$ at large $m$. However, in the effective model
the bifurcation is shifted to larger torque values. This might be
related to the ridge in the manifold of bifurcation points. Nevertheless,
considering the fact that we approximate the helical filament by a rigid 
rod whose helicity comes in through the friction coefficients, we obtain a
very good agreement with our simulations.

\section{Summary and conclusions}
\label{sec.cons}

Bacteria move forward by rotating a bundle of helical flagella which
creates a thrust force that pushes against the cell body. In this 
article we have modeled a single flagellum based on the discretized version
of Kirchhoff's elastic-rod theory and developed a coarse-grained approach
for driving the helical filament by a motor torque. When increasing the
motor torque, the thrust force reveals a supercritical Hopf bifurcation 
due to buckling of the helical filament. When the torque is further 
increased, a second buckling instability occurs. The Hopf bifurcation is 
also visible when we attach the flagellum to a spherical particle, which 
mimics the cell body, so that the whole model bacterium moves forward. 
Via the size of the cell body we can tune the thrust force pushing against 
the cell body and the critical torque for buckling changes. This results 
in a characteristic diagram critical force versus 
torque for the buckling transition [fig.\ \ref{fig: dynamic}(b)].

We have developed a theory for the observed buckling transition by 
approximating the helical filament by a helical rod with an effective 
bending rigidity and the characteristic rotation-translation coupling.
The basic picture is that the filament buckles under the frictional 
forces and torques that act along the filament when the filament rotates.
For large friction of the load particle, when its size is comparable
to a bacterial cell body, buckling is mostly due to the thrust force created 
along the filament and similar to a rod that buckles under its own weight. 
In the limit of small friction of the load particle, the critical thrust 
force tends to zero and buckling is mostly driven by the frictional torque 
acting along the 
filament. However, our modeling reveals that subtle details of the
specific problem are important. One has to take into account the precession 
of the buckled filament about the applied torque and, in particular, a 
perpendicular frictional force due to the motion of the model bacterium
that stabilizes the filament against buckling. Finally, taking into 
account the small 
deformation of the rotating helical filament, we are able to obtain a 
quantitative agreement with the simulated graph, critical force versus torque,
in the biologically relevant regime.

To further illustrate the biological relevance of the observed buckling
transition, we first summarize a few experimental values. Hotani gives 
the motor torque for observing a polymorphic transition of the flagellum
at around $1.1 \pico\newton\micro\meter$ \cite{Hotani1982}, whereas  
Darnton \emph{et al.} mention a mean torque acting on a flagellum of
about $1.4 \pico\newton\micro\meter$ \cite{Darnton2007a}. These values 
agree with the torques where we observe buckling for realistic cell body
sizes (see fig.\ \ref{fig: dynamic}). Reference \cite{Darnton2007a}
also mentions the relative stiffness of the helical filament so that
it hardly deforms under rotation which agrees with our simulations.
Finally, thrust forces created by the bundle are given as 
$0.41 \pm 0.23 \pico\newton$  \cite{Darnton2007a} or 
$0.5 \pico\newton$ \cite{Chattopadhyay2006}. This agrees with an estimate
$F=6\pi \eta a v \approx 0.6 \pico\newton$ where we take the radius of the 
load particle as $a=1\micro\meter$ and use the swimming velocity 
$v= 30 \micro\meter /\mathrm{s}$. All these values are close but below
the simulated values $F_{\mathrm{c1}} \approx 1 \pico\newton$ for real
cell sizes. However, we note that $F_{\mathrm{c1}}$ scales as 
$A_{\mathrm{eff}}/L_0^2$, as our analytic model shows, and thereby
depends on the explicit choice of the bending ($A$) and torsional ($C$) 
rigidities. We have chosen particular values for them and also set $A=C$ 
in our simulations. So $F_{\mathrm{c1}}$ will vary with the actual parameters.

It is clear that swimming bacteria should avoid buckling for efficient 
locomotion. However, they cannot simply increase bending rigidity $A$ 
since a certain flexibility is necessary during polymorphic transformations 
or when a bundle forms. Reference \cite{Turner2000} shows pictures where
single flagella are in a bent conformation similar to the buckled state 
in our simulations. This might be a hint that flagella naturally buckle 
under their own thrust. In peritrichous bacteria such as \textit{E.Coli} and 
\textit{Salmonella}, several flagella form a bundle which then has
larger bending stiffness and therefore buckling is not observed.

Monotrichous bacteria only use a single flagellum. Their conformation
differs in pitch and radius from the flagella of peritrichous 
bacteria \cite{Fujii2008}. A detailed analysis shows that their 
swimming efficiency is reduced due to a smaller pitch angle with
$\sin \alpha\approx 0.75$ \cite{Spagnolie2011}. This increases
the critical force $F_{cr} \propto 1/\sin\alpha$ by about $10\%$ 
compared to peritrichous bacteria and might be an adaption of the monotrichous 
bacteria to enhance the stability of their single flagellum. 

We also showed that a pulling flagellum is not stably aligned along the
applied torque. So most bacteria use their flagella to push themselves 
through the fluid. Nevertheless, there are some marine bacteria that
use a back-and-forth rather than a run-and-tumble strategy for chemotaxis. 
They live in a turbulent aqueous environment in the ocean where they 
experience large shear gradients on the micron scale \cite{Luchsinger1999}.
Simulations in Ref.\ \cite{Luchsinger1999} show that in addition to the 
shear-driven reorientation of the bacterium there must be further
contributions to the reorientation. Besides rotational diffusion this could 
also be the unstable orientation of the rotating filament when it pulls the
cell body.
Recent experiments on the back-and-forth motion of marine bacteria
\textit{Vibrio alginolyticus} directly show this reorientation of the flagellum
\cite{Xie2011}.

We close with this comment and hope that our work initiates a more
careful search for the buckling transition in bacterial flagella.


\begin{acknowledgement}
We thank M. Graham and R. Netz for stimulating discussions
and acknowledge financial support from the VW foundation within the
program "Computational Soft Matter and Biophysics" (grant no. I/83 942).
\end{acknowledgement}

\appendix

\section{Summary of resistive force theory for a helix} 
\label{app friction}

At low Reynolds number the force $\vec F$ and torque $\vec M$ acting on a 
particle of arbitrary shape are linearly related to its translational and 
rotational velocities \cite{Happel1983},
\begin{align}
 \begin{pmatrix}
  \vec F\\
  \vec M
 \end{pmatrix}
&= \begin{pmatrix}
 \mat A& \mat C\\
 \mat C^\mathtt{T} & \mat B
\end{pmatrix}
\begin{pmatrix}
  \vec v\\
  \vec \omega
\end{pmatrix}.
\label{eq: alg lin dyn in app}
\end{align}
The translational friction tensor $\mat A$, the rotational friction 
tensor $\mat B$, and the coupling tensor $\mat C$ are determined by the 
shape of the particle.
Note that
the rotational friction tensor 
$\mat B$ and the coupling tensor $\mat C$ depend on the choice of the origin of 
the coordinate system whereas the translational friction tensor $\mat A$ 
is unique.

In a moving helical filament, different parts interact via hydrodynamic
interactions. Nevertheless, using slender-body theory, Lighthill 
demonstrated that one can describe the hydrodynamic friction of the
filament with the help of resistive force theory
\cite{Lighthill1976,Childress1981}. In this theory one introduces local
friction coefficients per unit length parallel ($\gamma_\parallel$) and 
perpendicular ($\gamma_\perp$) to the tangent vector of the filament.
Lighthill adjusted the coefficients for the helical filament to
\cite{Lighthill1976}
\begin{align}
 \gamma_\parallel&= \frac{2 \pi \eta}{\ln(2q/r)} \enspace \mathrm{and} 
\enspace \gamma_\perp = \frac{4 \pi \eta}{\ln(2q/r)+1/2}.
\end{align}
Here $\eta$ is the shear viscosity, $r=0.02\micro\meter$ the cross-sectional 
radius of the bacterial flagellum, and $q$ a characteristic length, for which 
Lighthill derived $q=0.09\Lambda$, where $\Lambda = 2 \pi R / \cos\alpha$ is 
the filament length of one helical turn.

In a helical filament with translational velocity $\vec v$ and angular 
frequency $\vec \omega$ each segment moves with a velocity 
$\vec v + \vec \omega \times \vec r$, where $\vec r$ is the position vector
from a point on the helical axis to the segment. The force and torque 
densities to initiate such a motion are
\begin{align}
 \vec f & = \left(\gamma_\parallel \mat P_\parallel + \gamma_\perp \mat
 P_\perp \right) \cdot (\vec v + \vec \omega \times \vec r) \\
\vec m & = \vec r \times  \vec f,
\label{eq: app: local force}
\end{align}
where we use the projectors on the local tangent vector $\vec e_3$
and the space perpendicular to it,
\begin{align}
  \mat P_\parallel &= \vec e_3 \otimes \vec e_3, \quad \text{and}&
  \mat P_\perp &= \mat 1 - \vec e_3\otimes \vec e_3.
\end{align}
Integrating force and torque densities along the helical filament with 
position vector
\begin{align}\vec r(s) = \left(R \cos \left(\frac{\cos \alpha}{R}s\right), R 
\sin \left(\frac{ \cos\alpha}{R} s \right),\sin \alpha s\right)^\mathtt{T}
\label{eq: helix},
\end{align}
gives Eq.\ (\ref{eq: alg lin dyn in app}). For comparing theory and 
simulation in sect.\ \ref{sec: The motor-driven helical filament},
we calculated the integrals using the computational software program 
``Mathematica''. In particular, we took into account that the helical filament in the simulations does not consists of an integral number of helical turns and that the rotational axis is shifted against the helical axis.
In our analytical theory for the buckling transition in sect. 
\ref{sec.buckling theory}, we used friction coefficients calculated for
a full helical turn with $L = 2 \pi R /\cos \alpha$. The relevant
coefficients become
\begin{subequations}
\begin{align}
A_\perp  & = L\gamma_\perp 
\left(1+\frac{\delta}{2}\cos^2\alpha \right)
\label{eq: frict coeffiI}
 \\
A_\parallel & =L  \gamma_\perp 
\left(1+ \delta  \sin^2 \alpha \right)
\\
B_\parallel & = L \gamma_\perp  R^2 
\left(1+\delta(1 -\sin^2 \alpha)\right)
\\
C_\parallel  &= L \gamma_\perp R \delta  \sin \alpha
\cos \alpha ,
\end{align}
\label{eq: frict coeffi}
\end{subequations}
where we use $\delta=\frac{\gamma_\parallel-\gamma_\perp}{\gamma_\perp}$ to
characterize the anisotropy in the local friction coefficients.
Note that $A_\parallel$ also holds for arbitrary filament lengths when
$L$ is not an integer of a full helical turn. For all other coefficients 
one obtains corrections of the form 
$\sin \left(\frac{\cos \alpha }{R}L\right)$ that vanish in the limit
$L \to \infty$.

The effective friction coefficients for the helical rod used in 
Sec.\ \ref{sec.buckling theory} follow
by dividing the friction coefficients of Eqs. \eqref{eq: frict coeffi} 
by the rod length $L_0= L \sin\alpha$:
\begin{subequations}
\begin{align}
 a_\perp &= A_\perp / L_0, 
&
a_\parallel &= A_\parallel/L_0 
,\\
b_\parallel &= B_\parallel/L_0 
,
&
c_\parallel &= C_\parallel/L_0,
\label{eq: frict coeffi BC}
\end{align}
\end{subequations}

\section{Rotational motion of a rigid helix}
\label{app rotating helix}
%
%

\begin{figure}
\begin{center}
 \includegraphics[width=.8\columnwidth]{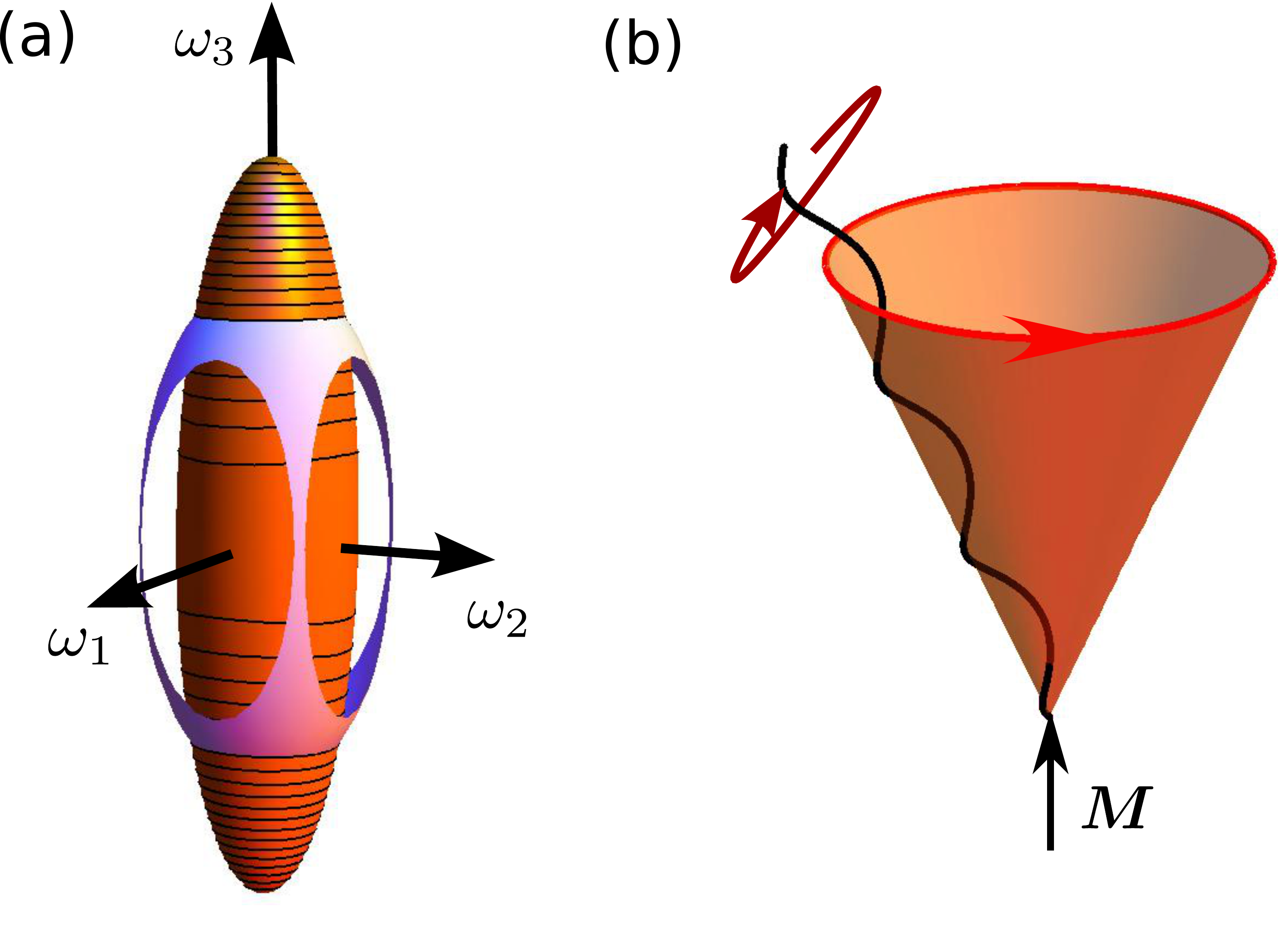}
\end{center}
\caption{(a) The constant applied torque and the dissipated energy
define two ellipsoids (red and blue) in the body fixed frame of a
rigid helix. The intersection gives the path of the angular velocity 
$\vec \omega$. For a long slender helix two directions are degenerate 
and the trajectories are circles. (b) In the lab frame the helix
rotates about its axis which precesses about the applied torque $\vec M$.
}
\label{fig.precession}
\end{figure}

Starting from Eq. \ (\ref{eq: alg lin dyn in app}) we set $\vec v=0$ and concentrate on the rotational motion due 
to a constant external torque $\vec M$ with the relevant equation
\begin{align}
\vec M&=\mat B  \vec \omega. \label{Glg: reine Rotation}
\end{align}
The rotational friction tensor $\mat B$ is symmetric. In the following
we use its frame of eigenvectors $\{\vec e_1,\vec e_2,\vec e_3\}$ and the 
eigenvalues $B_1$, $B_2$, and $B_3$. We differentiate Eq.\ 
(\ref{Glg: reine Rotation}) with respect to time $t$, use 
$\partial_t \vec e_i=\omega \times \vec e_i$, and obtain in the
eigenframe of $\mat B$,
\begin{align}
 B_1 \partial_t \omega_1 &= (B_2-B_3)\omega_2\omega_3,\\
 B_2 \partial_t \omega_2 &= (B_3-B_1)\omega_1\omega_3,\\
 B_3 \partial_t \omega_3 &= (B_1-B_2)\omega_1\omega_2.
\end{align}
These equations are the same as the Euler equations for a rigid body with
an inertia tensor $\mat B$ and without friction. The external torque
is zero so that angular momentum is conserved. Following this analogy 
and according to Eq.\ (\ref{Glg: reine Rotation}), the constant external
torque in our case corresponds to the angular momentum of the rigid body,
and the dissipated energy $P=\vec M \cdot \vec \omega$ to the rotational 
kinetic energy. Hence, besides the square of the applied torque 
$M^2 = \sum_i B_i^2 \omega_i^2$ also the dissipated energy 
$P = \sum_i B_i \omega_i^2$ is a conserved quantity and the trajectory
of $\vec \omega$ follows from the intersection of two ellipsoids as
illustrated in fig.\ \ref{fig.precession}(a). In particular, if two of the 
friction coefficients $B_i$ are equal, the angular velocity $\vec \omega$ 
precesses in real space on a cone about the direction of the torque 
\cite{Landau1976,Arnold1978}. 


We already calculated one component of the rotational friction tensor 
$\mat B$ of a helix with filament length $L$ in the previous section. In 
general, for a long slender helix like the normal form of the bacterial 
flagellum two eigenvalues of $\mat B$ are equal to a good approximation,
$B_1 \approx B_2 \propto L^3$. The third small friction coefficient 
$B_3\propto L$ belongs to the principal axis, which is parallel to the helical
axis, again to a good approximation. Hence, a rigid helical filament precesses
about the applied torque [fig.\ \ref{fig.precession}(b)] and does not 
align parallel to the torque as observed in our simulations.


\section{Effective bending rigidity of a helix}
\label{app A_eff}

We aim at replacing the helical filament by a rod with an effective bending 
rigidity $A_\text{eff}$. Our strategy is to apply a small constant torque 
perpendicular to the helical axis, rewrite the total elastic energy as a 
function of the torque, and compare this result with the case of a simple
rod to obtain $A_\text{eff}$.
To bend a simple rod with a constant curvature $\Omega$, one needs
the bending energy $\mathcal F= \frac{1}{2} A_\text{eff} \Omega^2 L_0$. 
Using the torque $M = A_\text{eff} \Omega$ [see, for example, 
Eq.\ \eqref{eq: torqu geometry cpl.}), we obtain
\begin{align}
 \mathcal F&=\frac{1}{2} \frac{M^2}{A_\text{eff}}L_0.\label{eq: A_eff rod}
\end{align}
Now we calculate the corresponding elastic energy for the helical filament. 
We apply a constant torque $\vec M = M \vec e_x$ perpendicular to the 
helical axis $\vec e_z$ and replace in Kirchhoff's energy density 
\eqref{Glg_freieEnergie1} the components of the angular strain 
vector $\vec \Omega$ by the components of the torque 
$\vec M = A \Omega_1 \vec e_1 + A (\Omega_2 - \kappa) \vec e_2 + 
C (\Omega_3 - \tau) \vec e_3$:
\begin{align}
f_\text{cl}& =\frac{1}{2} \left( \frac{M_1^2}{A}+\frac{M_2^2}{A}
+\frac{M_3^2}{C} \right).\label{eq: energy - torque}
\end{align}
Note that the components of the applied torque, 
$M_i = M  \vec e_x \cdot \vec e_i$, depend on the local material
frame $\{\vec e_1,\vec e_2,\vec e_3\}$ of the helical filament. In leading
order in $M$, we calculate the components $M_i$ for the undeformed
helical filament of Eq.\ \eqref{eq: helix} using the Frenet frame and 
integrate Eq.\ \eqref{eq: energy - torque} along the filament:
\begin{align}
  {\mathcal F} 
&= \frac{M^2 L}{4} \left[\frac{1}{A} + \frac{\sin^2 \alpha}{A} +\frac{\cos^2\alpha}{C} + O\left(\frac{\sin 2 k L}{2 k L}\right)\right],
\end{align}
where we used $k=\cos \alpha / R$. We compare this result with 
Eq.\ \eqref{eq: A_eff rod} and introduce the helix height $L_0= L \sin\alpha$
in order to identify the effective bending rigidity 
\begin{align}
 \frac{1}{A_\text{eff}}&= \frac{1}{2}\frac{1}{\sin \alpha}\frac{1}{ A} \left(1 + \sin^2 \alpha + \frac{A}{C} \cos^2\alpha\right).
\end{align}

\begin{figure}
\begin{center}
 \includegraphics[width=.9\columnwidth]{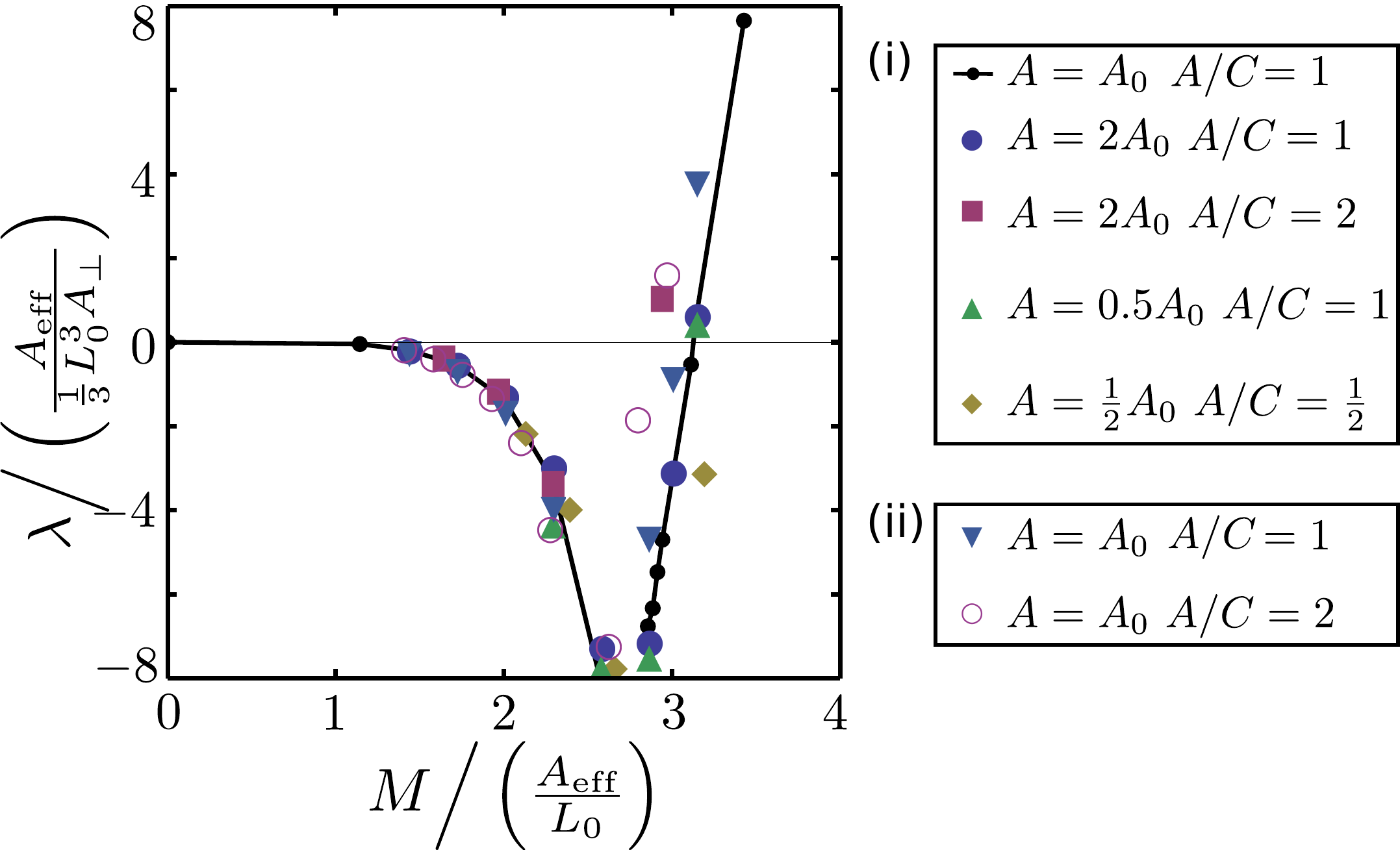}
\end{center}
\caption{Reorientation rate $\lambda$ as a function of motor torque $M$ 
for different elastic constants, where $A_0=3.5 \pico\newton \micro\meter^2$
is the bending rigidity used in this article. Two helical geometries of
the flagellum are considered: (i) for peritrichous bacteria used in this
article and (ii) for monotrichous bacteria.}
\label{fig: compare reorientation rate}
\end{figure}

To verify the applicability of the effective bending rigidity, we 
study in detail the reorientation rate $\lambda$ of the fixed flagellum,
when the thrust force pulls at it 
(see sect.\ \ref{sec: buckling - add. feat.}). Our claim is that the
reorientation rate depends on the bending of the helical filament
as a whole and thus $A_{\text{eff}}$ should be the relevant parameter.
We therefore determined $\lambda$ as a function of the motor torque $M$
for different values of the bending rigidity $A$ and the torsional 
rigidity $C$ (see fig.\ \ref{fig: compare reorientation rate}). 
In addition to the helical geometry of peritrichous bacteria 
used in this paper, we also considered the flagellum of monotrichous 
bacteria, which has different helical parameters: 
$\kappa_0=2.2\per\micro\meter$ and $\tau_0=2.5\per\micro\meter$ 
\cite{Fujii2008}. Dimensional analysis suggests to rescale the torque $M$ 
by the characteristic bending moment $A_\text{eff}/L_0$ as in 
Sect. \ref{sec.buckling theory} and the reorientation rate $\lambda$ 
by $A_\text{eff}/(\frac{1}{3} L_0^3A_{\perp})$, where $A_{\perp}$ is the friction 
coefficient introduced in Eq.\ (\ref{eq: frict coeffiI}). With such a 
rescaling all different curves for the reorientation rate fall onto a
common master curve in fig. \ref{fig: compare reorientation rate}. The 
effective bending rigidity $A_\text{eff}$ is therefore the right 
parameter in an effective description of the helical filament.



%


%

\end{document}